\documentclass[a4,aps,showpacs,amsmath,floatfix,twocolumn]{revtex4}
\usepackage{amsmath, bm}
\usepackage{graphicx}
\usepackage{dcolumn}
\usepackage{dirtytalk}
\usepackage{color}
\usepackage{amsmath}
\usepackage{here}
\usepackage[utf8x]{inputenc}
\usepackage{bm}
\usepackage[mathscr]{euscript}
\usepackage{amssymb}
\usepackage{amsmath}

\usepackage{hyperref}
\hypersetup{breaklinks=true,
            colorlinks=false,
            pdfborder={0 0 0}}
\usepackage{bm}
\usepackage{braket}
\usepackage{MnSymbol}
\usepackage{graphicx}
\graphicspath{{figures/}}
\usepackage{dsfont}

\usepackage{titlesec}

\begin{document}
\title{Rabi Regime of Current Rectification in Solids}

\author{Oles Matsyshyn$^{1}$, Francesco Piazza$^{1}$, Roderich Moessner$^{1}$, Inti Sodemann$^{1,2,3}$}

\affiliation{$^{1}$Max Planck Institute for the Physics of Complex Systems, Dresden 01187, Germany
}
\affiliation{$^{2}$Department of Physics and Astronomy, University of California, Irvine, California 92697, USA }
\affiliation{$^{3}$Institut f{\"u}r Theoretische Physik, Universit{\"a}t Leipzig, D-04103, Leipzig, Germany}

\begin{abstract}
We investigate rectified currents in response to oscillating electric fields in systems lacking inversion and time-reversal symmetries. These currents, in second-order perturbation theory, are inversely proportional to the relaxation rate, and, therefore, naively diverge in the ideal clean limit. Employing a combination of the non-equilibrium Green function technique and Floquet theory, we show that this is an artifact of perturbation theory, and that there is a well-defined periodic steady-state akin to Rabi oscillations leading to finite rectified currents in the limit of weak coupling to a thermal bath. In this Rabi regime the rectified current scales as the square root of the radiation intensity, in contrast with the linear scaling of the perturbative regime, allowing to readily diagnose it in experiments. More generally, our description provides a smooth interpolation from the ideal Periodic Gibbs Ensemble describing the Rabi oscillations of a closed system to the perturbative regime of rapid relaxation due to strong coupling to a thermal bath.
\end{abstract}


\maketitle

\textit{\color{blue} Introduction}. Crystalline solids lacking inversion symmetry can display bulk photovoltaic effects (BPVE)~\cite{Belinicher1980,BaltzN,Sturman1992}, namely macroscopic DC rectified currents in response to spatially uniform AC electric fields. There is a long tradition of studying these BPVE \cite{PhysRevB.19.1548,PhysRevB.23.5590,Belinicher,PhysRevB.52.14636,PhysRevB.61.5337,Sturman2020}, but also a growing renewed interest in investigating their connections to the Berry’s phase geometry and topology of electronic bands \cite{PhysRevLett.109.116601,PhysRevLett.115.216806,Morimotoe1501524,NaMo,PhysRevLett.123.246602,PhysRevB.99.045121,deJuan2017,matsyshyn2020berry,PhysRevB.95.041104,vanderbilt2018berry,PhysRevLett.105.026805,Kang2019,Ma2019}, and their potential for novel photovoltaic technologies \cite{PhysRevLett.109.116601,BrehmYoung,PhysRevLett.119.067402,Cook2017,PhysRevLett.121.267401,matsyshyn2020berry,Kumar2021}.

Our study is motivated by the following question: what is the ultimate fate of current rectification in Bloch bands in the ideal 
limit where relaxation times become very large? As we will demonstrate, there is in fact a well defined periodic steady state in such a limit, that we will refer to as the “Rabi regime”, in which the system sustains a finite DC rectified current.  

A useful starting point to appreciate the non-trivialities of such a clean limit is to analyze the problem perturbatively in the amplitude of electric field, as commonly done in most studies (see however Ref.\cite{2020arXiv200903596K,PhysRevB.52.2090,Morimotoe1501524,Leppenen2019,dantas2020nonperturbative}). Perturbation theory predicts a rectified current $\mathbf{j}$, that grows as the square of the amplitude of the field, $\mathbf{j} \propto |\mathbf{E}|^2$. For frequencies above the threshold for inter-band transitions, such perturbative BPVE are often separated into two mechanisms known as the shift and the injection current effects \cite{PhysRevB.19.1548,PhysRevB.23.5590,Sturman1992,PhysRevB.52.14636,PhysRevB.61.5337,Morimotoe1501524,NaMo,PhysRevB.99.045121,PhysRevLett.123.246602,matsyshyn2020berry,Belinicher,Sturman2020,BrehmYoung,PhysRevLett.109.116601,PhysRevLett.119.067402,Cook2017,PhysRevLett.121.267401,PhysRevB.52.2090,Leppenen2019,dantas2020nonperturbative,deJuan2017}. The injection current originates from difference of the band-diagonal velocity of the empty and occupied bands at a given crystal momentum $\mathbf{k}$. The shift current, on the other hand, originates from the difference of positions of Bloch wave-functions between the empty and occupied bands at a given $\mathbf{k}$, and can be computed as the contribution arising from the band-off-diagonal velocity operator. 

\begin{figure}[t]
\includegraphics[width=0.95\linewidth]{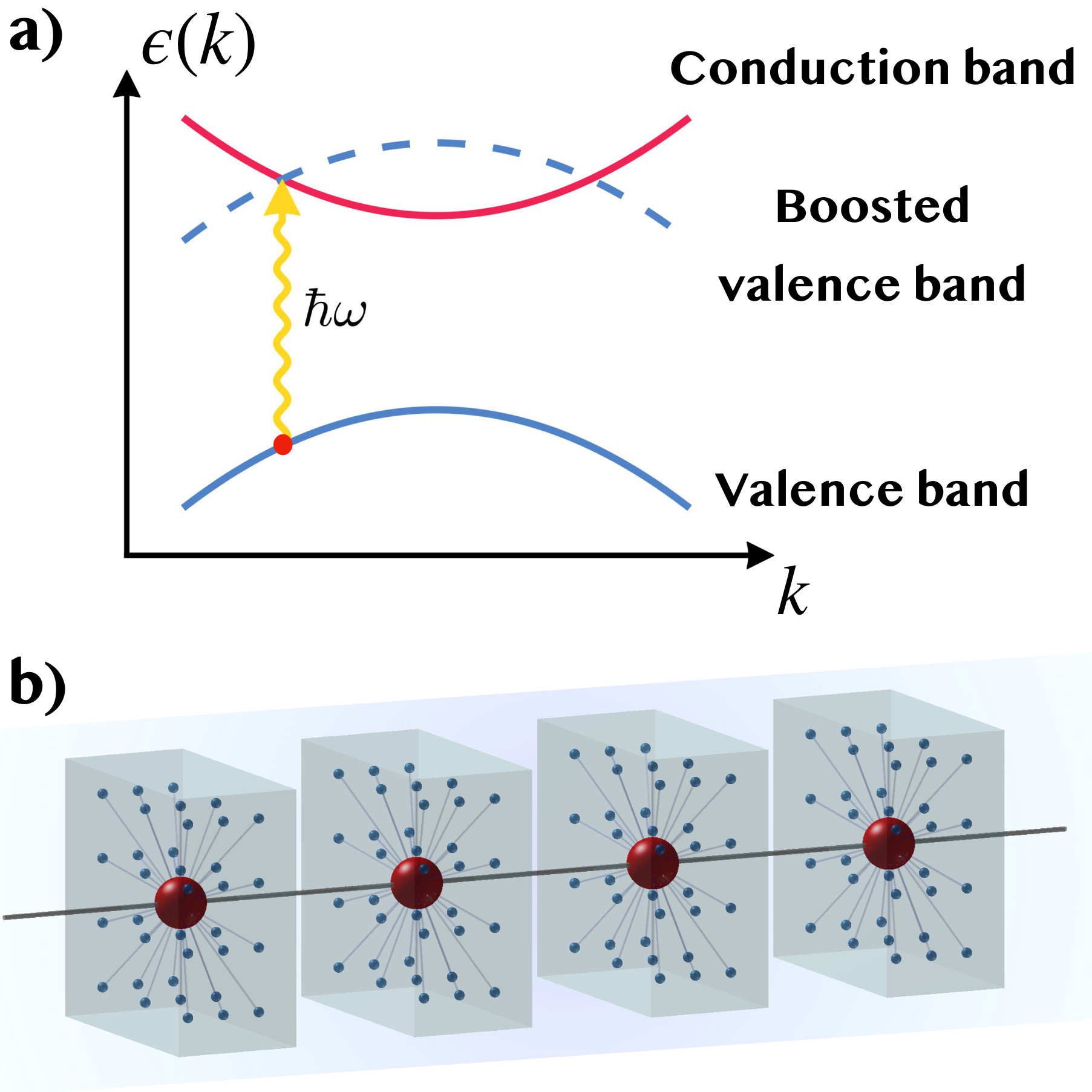}
  \caption{a) Energy crossing between boosted valence and conduction bands in Floquet representation. b) Depiction of underlying tight binding model with physical sites (red balls) which are tunnel coupled (solid lines) among themselves and with their own identical fermionic bath (blue balls).}
  \label{fig:fig1}
\end{figure}

A crucial distinction between the shift and injection currents is that, within perturbation theory, the shift current appears to have a finite value in the “clean” limit of vanishing relaxation rate, $\Gamma\rightarrow 0$, while the injection current appears to diverge in such limit as $1/\Gamma$, \textcolor{black}{which ultimately arises from the vanishing quasi-energy denominators appearing at higher orders of the perturbation theory for the rectified current} (see Refs.\cite{PhysRevB.19.1548,PhysRevB.23.5590,Sturman1992,PhysRevB.52.14636,PhysRevB.61.5337,Morimotoe1501524,NaMo,PhysRevB.99.045121,PhysRevLett.123.246602,matsyshyn2020berry,Belinicher,Sturman2020,BrehmYoung,PhysRevLett.109.116601,PhysRevLett.119.067402,Cook2017,PhysRevLett.121.267401,PhysRevB.52.2090,Leppenen2019,dantas2020nonperturbative,deJuan2017} and S.I.I A). Such divergence is often handled in an ad-hoc manner by computing the response of the rate of change of the current, $d\mathbf{j}/dt$, and assuming that such growth leads to a current saturation to a value proportional to the relaxation time $\tau\sim \hbar/\Gamma$. However, recently an interesting non-perturbative study of the CPGE in Weyl semimetals \cite{Leppenen2019} demonstrated that the rectified current saturates to a finite value even in the clean limit of vanishing relaxations ($\Gamma \rightarrow 0$) within a semiclassical kinetic framework. The underlying mechanism for such saturation is the Rabi dynamic broadening of absorption~\cite{PhysRevLett.42.1495,PhysRevB.21.3502,Parshin,Parshin2}, which occurs when the energy scale controlling the transitions between conduction ($c$) and valence ($v$) bands exceeds the relaxation rate, $ e\mathbf{E}\cdot\bra{c}\mathbf{r}\ket{v}\gg\Gamma$, which we refer to as the Rabi regime. 

In the present work we develop a microscopic description of the currents for arbitrary values of the non-linearity parameter $e\mathbf{E}\cdot\bra{c}\mathbf{r}\ket{v}/ \Gamma$, that captures the perturbative and the Rabi regimes on an equal footing. To do so, we employ a Keldysh-Floquet formalism~\cite{KelFloq6,PhysRevB.50.5528,kamenev_2011} in a generic two-band system coupled to an ideal fermionic bath, following the pioneering approach of Ref.\cite{Morimotoe1501524} (see also Ref.\cite{2020arXiv200903596K}). As we will see, and contrary to the expectations of perturbation theory, in the Rabi regime, the traditional resonant shift current contributions vanish, whereas the injection currents approach a finite limit that scales as the square root of the radiation intensity in a sharp contrast to the perturbation theory expectations. We will also demonstrate that the Rabi regime can be viewed as an example of thermalizing synchronization of a system under an external periodic drive that can be described by the periodic Gibbs ensemble~\cite{Moess4,PhysRevLett.109.257201}.

\textit{\color{blue} Keldysh-Floquet Formalism}. We derive the non-perturbative expression for currents within a two-band model (see Fig.\ref{fig:fig1}(a)). The electric current operator is defined as: $\hat{\mathbf{j}} = e \hat{\mathbf{v}}/\hbar= \partial \hat{H}_0(\mathbf{k}+e\mathbf{A}(t)/\hbar)/\partial \mathbf{A}(t)$, where $\hat{H}_0(\mathbf{k})$ is the 2x2 matrix Bloch Hamiltonian, and $\mathbf{A}(t)$ is the vector potential from spatially uniform but time dependent electric field. Since the crystal momentum $\mathbf{k}$ is conserved, the problem is equivalent to a collection of independent driven two-level systems.
We restrict our analysis to a monochromatic electric field with frequency $\omega$:
\begin{equation}\label{efdef}
     \mathbf{A}_0(t) = i\frac{\mathbf{E}}{\omega}e^{i\omega t}-i\frac{\mathbf{E}^{*}}{\omega}e^{-i\omega t}.
\end{equation}
\textcolor{black}{Here ${\bf E}$ is a vector with complex entries, allowing us to capture light of arbitrary degree of polarization, including the case of linear polarization, when all components can be chosen to be real, to fully circularly polarized light, when two orthogonal components differ by a phase of $\pi/2$.} The periodicity in time allows to employ the Floquet picture (for details see S.I.I B) where multiple Floquet bands appear with a quasi-energy that is boosted by multiples of the driving frequency (see FIG.\ref{fig:fig1}(a)). We simplify the problem by truncating the Floquet Hamiltonian to two bands that are in resonance, in the spirit of a rotating wave-approximation \cite{NaMo}. \textcolor{black}{This approximation is well justified when the off-diagonal terms in the Floquet Hamiltonian are smaller in comparison to the Floquet quasi-energy difference to other remote Floquet bands, namely when $e|\mathbf{E}\cdot \bra{c}\mathbf{r}\ket{v}\ll\hbar\omega$ (see e.g. Ref.\cite{PhysRevB.88.155129} and S.I.I B).} Thus the approximate Floquet Hamiltonian is:
\begin{equation}\label{FlHam}
    H_\mathbf{F} = 
    \left(\begin{array}{cc}\epsilon_1+\hbar\omega&i\frac{e\mathbf{E}}{\hbar\omega}\cdot \left(\frac{\partial H_0(\mathbf{k})}{\partial \mathbf{k}}\right)_{12}\\ 
    -i\frac{e\mathbf{E}^*}{\hbar\omega}\cdot \left(\frac{\partial H_0(\mathbf{k})}{\partial \mathbf{k}}\right)_{21}&\epsilon_2\end{array}\right) = h_0+\mathbf{h}\cdot\boldsymbol{\sigma},
\end{equation}
where 1 stands for valence, 2 for conduction and $\epsilon_{1,2}$ are effective valence and conduction band energies respectively (which could be dressed by higher order perturbative corrections with respect to bare band energies, as further discussed in S.I.I B). The subscript $\mathbf{F}$ stands for the representation of the operator in the Floquet picture, which is related to the ordinary Schrödinger picture as follows:
\begin{equation}\label{FStr}
    \hat{O}_\mathbf{F} = \left(\begin{array}{cc}
        O_{22} & O_{12}  \\
        O_{21} & O_{11}
    \end{array}\right), \quad   \hat{O}(t) = \left(\begin{array}{cc}
        O_{11} & O_{21}e^{-i\omega t}  \\
        O_{12}e^{i\omega t} & O_{22}
    \end{array}\right).
\end{equation}

In order to capture relaxation processes, we couple the system to a bath and apply the non-equilibrium Green function technique on the Keldysh contour (see S.I.I C and Refs.\cite{NaMo,KelFloq1,KelFloq2,KelFloq3,KelFloq4,KelFloq5,KelFloq6}). We choose a simple model in which each fermionic site in the system of interest is coupled to its own fermionic bath, with a common hopping amplitude $V_{\mathrm{mix}}$ (see Fig.\ref{fig:fig1}(b)). The temperature of the bath is $T_{\text{bath}} = 1/(k_B\beta)$ and the chemical potential is $\mu$. 
The effective density matrix of the system is given by the lesser equal time Green Function $G^<(t,t)$, and can be shown to be (see S.I.I C):
\begin{multline}\label{lGF}
    \hat{\rho}_\mathbf{F} = -i\hat{G}^<_\mathbf{F} = \left(\begin{array}{cc}
         f_1&0  \\
         0&f_2 
    \end{array}\right)+\\+\frac{f_1-f_2}{2(h^2+\frac{\Gamma^2}{4})}\left(\begin{array}{cc}
         -h_x^2-h_y^2& h_-(h_z+i\frac\Gamma2) \\\\h_+(h_z-i\frac\Gamma2)
         &h_x^2+h_y^2 
    \end{array}\right),
\end{multline}
where $f_{1,2} = 1/(1+\exp(-\beta(\epsilon_{1,2}-\mu)))$ are valence and conduction Fermi-Dirac occupation factors respectively, $h_\pm = h_x\pm ih_y$ and $\Gamma = 2\pi |V_{\mathrm{mix}}|^2$ is the relaxation rate. 

The DC current of the system, ${J}^\alpha = -ie\mathbf{Tr}\left[\hat{G}^<_{\mathbf{F}}\hat{v}^\alpha_{\mathbf{F}}\right]/\hbar$, can be decomposed into three contributions:
\begin{gather}
    J_1^\alpha = \frac{e}{\hbar}\int\frac{d\mathbf{k}}{(2\pi)^3}\frac{\frac{\Gamma}{2}(f_{1}-f_2)}{h^2+\Gamma^2/4}(h_yv^\alpha_x-v^\alpha_yh_x),\label{resshift}\\
    J_2^\alpha = \frac{e}{\hbar}\int\frac{d\mathbf{k}}{(2\pi)^3}\frac{h_z(f_{1}-f_2)}{h^2+\Gamma^2/4}(h_xv_x^\alpha+h_yv_y^\alpha),\label{nonreshift}\\
    J_3^\alpha = -\frac{e}{\hbar}\int\frac{d\mathbf{k}}{(2\pi)^3}(f_1-f_2)v^\alpha_z\frac{h_x^2+h_y^2}{h^2+\frac{\Gamma^2}{4}}.\label{inject}
\end{gather}

Here $\alpha$ denotes the real space indices and the velocity operator in Floquet representation $\hat{v}^\alpha_{\mathbf{F}}$ is decomposed in the Pauli basis, namely $\hat{v}^\alpha_{\mathbf{F}} = \sum_{i=x,y,z}v^\alpha_i\sigma_i$. In the supplementary (see S.I.I E) we compare the currents Eq.(\ref{resshift}-\ref{inject}) with perturbation theory and show that Eq.(\ref{resshift}) and Eq.(\ref{inject}) recover the resonant behaviour of the shift and injection currents respectively, whereas Eq.(\ref{nonreshift}) becomes the non-resonant component of the shift current in the limit $e\mathbf{E}\bra{c}\mathbf{r}\ket{v}\ll\Gamma$.

Now, to analyse the clean limit behaviour of the injection current Eq.(\ref{inject}), it is useful to take the approximation in which both the diagonal $h_z$ matrix elements in Eq.(\ref{FlHam}) are greater than $\Gamma$ and the off-diagonal elements $h_{x,y}$. This is typically well satisfied in most solids except in special situations such as resonant absorption on extremely flat bands, and we will demonstrate that this is a good approximation by explicit calculations later on. Therefore we replace $1/(h^2+\Gamma^2/4)\approx \pi\delta(h_z)/\sqrt{h_x^2+h_y^2+\Gamma^2/4}$, which leads to the following expression:
\begin{multline}\label{InjCur}
    {J}^\alpha_{3}=\pi  \frac{e}{\hbar}\int\frac{d\mathbf{k}}{(2\pi)^3}(f_2-f_1)(v^\alpha_1-v^\alpha_2)\times\\\times\frac{\left|\frac{e\mathbf{E}}{\hbar\omega}\cdot \left(\frac{\partial H_0(\mathbf{k})}{\partial \mathbf{k}}\right)_{12}\right|^2}{\sqrt{\left|\frac{e\mathbf{E}}{\hbar\omega}\cdot \left(\frac{\partial H_0(\mathbf{k})}{\partial \mathbf{k}}\right)_{12}\right|^2+\frac{\Gamma^2}{4}}}\delta(\epsilon_1-\epsilon_2 +\hbar\omega),
\end{multline}
where $\mathbf{v}_{1,2}$ are conduction  and valence band velocities respectively. We again see that in the limit of fast relaxation ($e\mathbf{E}\cdot\bra{c}\mathbf{r}\ket{v}\ll \Gamma $) Eq.(\ref{InjCur}) reproduces the behaviour predicted by perturbation theory (see S.I.I A). Remarkably, however, in the clean limit ($\Gamma \rightarrow 0$), the above formula predicts a finite current, in sharp constrast to the naive extrapolation of perturbative result. In other words, the relaxation rate in the denominator of the perturbative expressions acquires a non-perturbative modification by the driving electric field of the form:
\begin{equation}
    \frac{1}{\Gamma}\rightarrow \frac{1}{\sqrt{4\left|\frac{e\mathbf{E}}{\hbar\omega}\cdot \left(\frac{\partial H_0(\mathbf{k})}{\partial \mathbf{k}}\right)_{12}\right|^2+\Gamma^2}}
\end{equation}
\begin{figure*}
\centering
\includegraphics[width=0.98\textwidth]{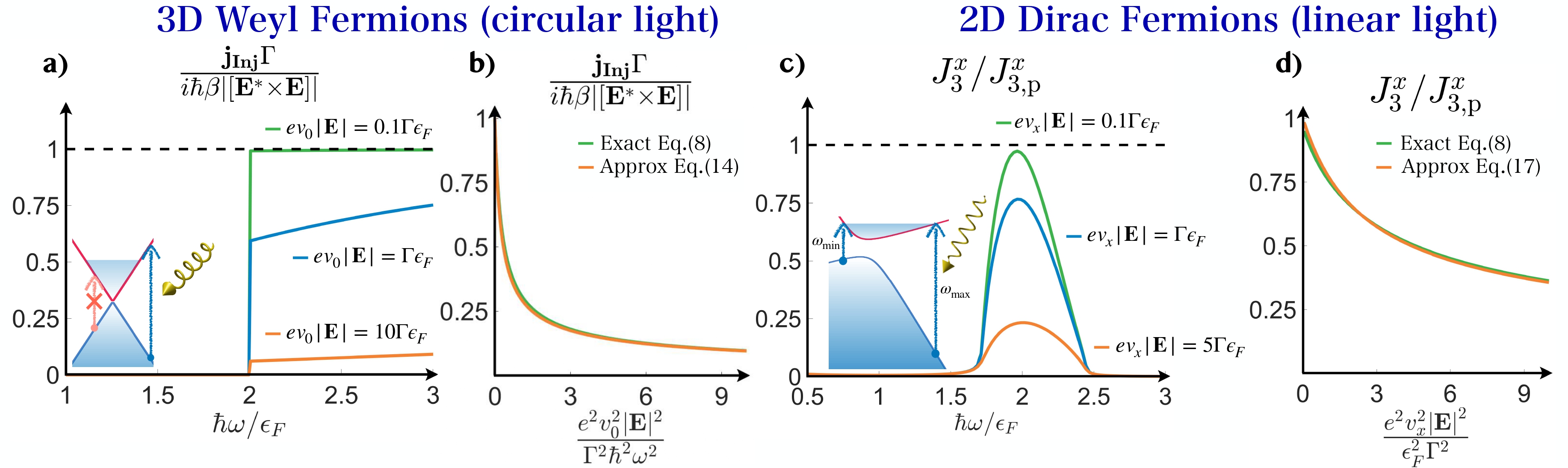}
\caption{a) Rectified current dependence on frequency, and (b) on electric field amplitude for 3D Weyl fermion ($\mathbf{E}/|\mathbf{E}|=(0,i\sin\pi/8\sin\pi/4,\cos\pi/4+i\cos\pi/8\sin\pi/4)$), c) Rectified current dependence on frequency, and (d) on electric field amplitude for 2D Dirac fermion ($\mathbf{E}/|\mathbf{E}|=(1,0),u_x/v_x = 0.2, m = 0.5\epsilon_F$).}
\label{fig:fig3}
\end{figure*}

Therefore in the clean limit, the injection current scales as the absolute value of the electric field, $\mathbf{J}_{3}\propto|\mathbf{E}|$, and, accordingly, it is proportional to the square root of the radiation intensity.  On the other hand, the term $\mathbf{J}_1$ from Eq.(\ref{resshift}), which reduces to the usual resonant shift current from perturbation theory (see S.I.I A for details), can be seen to vanish in the clean limit $\Gamma\rightarrow0$ from Eq.(\ref{resshift}). This is noteworthy because in the perturbative regime ($ e\mathbf{E}\bra{c}\mathbf{r}\ket{v}\ll\Gamma$ ) the shift current naively approaches a finite value in the $\Gamma\rightarrow0$ limit.

\textit{\color{blue} Synchronization and Rabi Limit of Rectification}. While the Keldysh formalism allows for a description with arbitrary strength of coupling to the bath, there is a simpler way to understand the ideal behavior in the limit of vanishing coupling to the bath ($\Gamma \rightarrow 0$). In fact, this limit can be understood simply as a form of Rabi oscillations associated with the inter-band transitions driven by the oscillating field. We will describe how to understand this limit within the picture of the Periodic Gibbs Ensemble (PGE) \cite{Moess4,Moess1,Moess2,Moess3} that captures the steady state synchronization of the system with the driving field.

Consider an initial state described by a density matrix $\rho_0$. This density matrix can be decomposed in the eigenstates of the time dependent Hamiltonian, $\psi_{\alpha}(t)$, and therefore the state at any later time $t$, is given by:
\begin{equation}
    \rho(t) = \sum_{\alpha\beta}\rho_{\alpha\beta}\psi_\alpha(t)\psi^\dagger_\beta(t),
\end{equation}
where $\rho_{\alpha\beta} = \mathrm{Tr}\left[\rho_0 \psi_{\alpha}(t_0)\psi^\dagger_\beta(t_0)\right]$. Now, the Floquet theorem implies that, barring accidental degeneracies, the operators $\psi_\alpha(t) \psi_\beta(t)^\dagger$ are only periodic when $\alpha = \beta$. The late-time synchronization associated with the PGE can be understood as a process in which the memory of these off-diagonal amplitudes of the density matrix in the Floquet basis disappears in a kind of thermalization process, leading to a steady state that is exactly periodic and synchronized with the drive (see S.I.I G):
\begin{equation}
    \rho_{\mathbf{PGE}} = \sum_\alpha \rho_{\alpha\alpha}\psi_\alpha(t)\psi^\dagger_\alpha(t).
\end{equation}

Remarkably the above ensemble is identical to the one that we have obtained within the Keldysh formalism in the limit of $\Gamma\rightarrow 0$, when one chooses the initial state $\rho_0$ to be the equilibrium Fermi-Dirac density matrix in the absence of the periodic perturbation, with the chemical potential and the temperature of the bath (see S.I.I E. for more details). In fact, within the same rotating-wave approximation used to solve the Floquet problem, this density matrix is explicitly given by (see S.I.I G):
\begin{multline}\label{PGEmaint}
    \hat{\rho}_{\mathbf{PGE}}(t)= \left(\begin{array}{cc}
         f_2&0  \\
         0& f_1
    \end{array}\right)+\\+\frac{f_1-f_2}{2h^2}\left(\begin{array}{ccc}
         h_x^2+h_y^2&h_z h_+ e^{-i\omega t}  \\\\h_zh_-e^{i\omega t}
         & -h_x^2-h_y^2
    \end{array}\right).
\end{multline}

This density matrix encodes the physics of Rabi oscillations (see S.I.I F for details). The above reduces exactly to the density matrix in Eq.(\ref{lGF}) in the clean limit $\Gamma\rightarrow0$, once it is expressed in the Floquet picture (see Eq.(\ref{FStr})), and therefore predicts the same rectification currents that we have previously described in the clean limit.

We would like to note that most studies of PGE to date have focused on what might be called “internal” synchronization, which considers a closed system acting as its own bath. In this context, the initial condition, $\rho_0$, is freely chosen and it is not unique. In our context, however, the emergence of the PGE follows from different principles. Coupled to the bath, the system loses memory of its initial state at late times. It does so by flowing towards a \textit{unique} stable periodic solution. Remarkably, in the limit of weak coupling to the bath, this steady state coincides exactly with one specifically chosen PGE, whose initial condition is the one associated with the thermal equilibrium system with an infinitesimal coupling to the bath in the absence of the periodic drive.

\textcolor{black}{Therefore, although we have performed our calculations in a rather specific microscopic setting, we have been able to recover the universality of the PGE in the limit of weak coupling to the bath that we are using. Since the PGE can be justified under generalized entropy maximixation principles \cite{Moess4,Moess1,Moess2,Moess3}, this is a compelling indication that our results describe the behavior of a large class of systems coupled to ideal heat baths.}

\textit{\color{blue} Photocurrents for 3D Weyl and 2D Dirac Fermions.} One important distinction between shift and injection currents is their transformations under time-reversal (TR) symmetry~\cite{ahn2020lowfrequency,hornung2020quantum}. The shift current can exist in TR invariant materials illuminated with linearly polarized light, whereas the injection current requires breaking of TR symmetry, namely either by shining linearly polarized light on a TR broken material~\cite{1978JETPL..27..604I,Zhang2019}, or by shining circularly polarized light, also known as the circular-photogalvanic-effect (CPGE), which has an interesting manifestation in Weyl semimetals~\cite{PhysRevB.95.041104,deJuan2017,PhysRevB.98.075305,PhysRevB.98.155145,conv,PhysRevLett.124.196603,Rees2020,ni2020giant,dantas2020nonperturbative,Ni2021,Ma2017,Nagaosa2020,KaiSun117203,Ji2019}.

We would like to illustrate this behavior for representative nodal fermions with Hamiltonians that are linear in momentum. These linear in $\mathbf{k}$ Hamiltonians have negligible shift currents (see details in S.I.I J) and therefore allow us to focus on the behavior of injection currents, which we will consider from here on in this section. We will consider two types of model that are relevant to a large class of materials. The first is an ideal 3D Weyl fermion, and our focus will be on the non-perturbative modifications to the CPGE. As we will see, our results are in perfect agreement with those obtained recently in Ref.\cite{Leppenen2019}. The second will be a 2D tilted Dirac massive fermion, and our focus will be to investigate the non-perturbative regime of rectification for linearly polarized light in a time reversal breaking system. 

The ideal 3D Weyl Hamiltonian is: 
\begin{equation}\label{3DW}
    \hat{H}_0 = v_0\sum_{\alpha=x,y,z}{k_\alpha}\cdot{\hat{\sigma}_\alpha}.
\end{equation}
Here $v_0$ is a Fermi velocity and $\hat{\sigma}_\alpha$ are Pauli matrices. This model respects TR but breaks inversion symmetry. When the system has a finite chemical potential, light absorption occurs above a threshold frequency $\hbar\omega>2\epsilon_F$ (see Fig.\ref{fig:fig3}(a)). By using the formula from Eq.(\ref{InjCur}), one obtains the following non-perturbative approximate expression of the injection current above such a threshold (see S.I.I I for details):
\begin{multline}\label{fres}
    \mathbf{J}_3 \approx \frac{i\pi^2 e^2 \omega}{v_0(2\pi)^3} \frac{ \left[\mathbf{E}^*\times\mathbf{E}\right]}{60\sqrt{|\mathbf{E}|^2+\frac{\Gamma^2\hbar^2\omega^2}{4v_0^2e^2}}} \frac{12|\mathbf{E}|^2+5\frac{\Gamma^2\hbar^2\omega^2}{v_0^2e^2} }{|\mathbf{E}|^2+\frac{\Gamma^2\hbar^2\omega^2}{4v_0^2e^2}},
\end{multline}
\textcolor{black}{where $|\bf{E}|^2={\bf E}^* \cdot {\bf E}$, with ${\bf E}$ understood as the complex vector defined in Eq.(\ref{efdef})  (see S.I.I I for comparison of this approximate formula against direct evaluation from the integral in Eq.(\ref{InjCur})). Eq.(\ref{fres})} in the perturbative regime ($e^2v_0^2|\mathbf{E}|^2 \ll \Gamma^2\hbar^2\omega^2$) approaches the known result~\cite{PhysRevB.95.041104,deJuan2017}
$\mathbf{J}_3\approx\hbar\beta\left[\mathbf{E}^*\times\mathbf{E}\right]/\Gamma,$
where $\beta = i\pi e^3/(3h^2)$. However, in the Rabi regime ($e^2v_0^2|\mathbf{E}|^2 \gg \Gamma^2\hbar^2\omega^2$) interestingly, the injection current approaches a value that is independent of the relaxation rate and it is given by:
\begin{equation}\label{CPGEol}
    \mathbf{J}_3\approx \zeta\frac{\beta\hbar^2\omega}{ev_0|\mathbf{E}|}\left[\mathbf{E}^*\times\mathbf{E}\right].
\end{equation}
Here $\zeta \approx 0.3$ is a numerical pre-factor with a weak dependence on the degree of light polarization. Its value for perfectly circularly-polarized light can be computed exactly from Eq.(\ref{InjCur}) to be $\zeta=1/(2\sqrt{2})$, in agreement with Ref.\cite{Leppenen2019} (see S.I.I I  for details).
The behavior of the rectified current in these two regimes and their crossovers are shown in Fig.\ref{fig:fig3}(a,b). 

We will now consider a 2D Dirac Hamiltonian given by:
\begin{equation}
    \hat{H} = u_x k_x \hat{\mathds{1}} + v_xk_x\hat{\sigma}_x+v_yk_y\hat{\sigma}_y +m\hat{\sigma}_z,
\end{equation}
where $m$ is the mass which breaks time-reversal symmetry, $v_x,v_y$ are anisotropic Fermi velocities, and $u_x>0$ is the tilt term that breaks inversion. The above model features absorbtion within a window of frequency given by $(2\epsilon_F-2\alpha\sqrt{\epsilon_F^2-m^2+\alpha^2m^2})/(1-\alpha^2)<\hbar\omega<(2\epsilon_F+2\alpha\sqrt{\epsilon_F^2-m^2+\alpha^2m^2})/(1-\alpha^2)$ (see Fig.\ref{fig:fig3}(c)). In this window the maximum current occurs when $\hbar\omega\approx 2 \epsilon_F$ (see Fig.\ref{fig:fig3}(c)), and for the electric field along the tilt direction. From Eq.(\ref{InjCur}), the corresponding component of the injection current can be approximated as (see S.I.I I):
\begin{multline}
    J^x_{3,\mathrm{max}}\approx \frac{e^2v_x}{\hbar v_y}\frac{\sqrt{\epsilon_F^2-m^2}}{60\pi}\frac{\frac{|\mathbf{E}|^2}{\epsilon_F^2}}{\sqrt{\frac{|\mathbf{E}|^2}{\epsilon_F^2}+\frac{\Gamma^2}{e^2v_x^2}}}\times\\\times\frac{5\frac{\Gamma^2}{e^2v^2_x}(1+2\frac{m^2}{\epsilon_F^2})+\frac{|\mathbf{E}|^2}{\epsilon_F^2}(6+13\frac{m^2}{\epsilon_F^2}-4\frac{m^4}{\epsilon_F^4})}{\frac{|\mathbf{E}|^2}{\epsilon_F^2}+\frac{\Gamma^2}{e^2v_x^2}}.
\end{multline}
Within perturbation theory this current would be $J_{3,\text{p}}^x =e^3v_x^2\sqrt{\epsilon_F^2-m^2}(1+2m^2/\epsilon_F^2)|\mathbf{E}|^2/(12\pi\hbar v_y\Gamma\epsilon_F^2)$. We use $J_{3,\text{p}}^x$ to normalise the numerical non-pertubative results shown in Figs.\ref{fig:fig3}(c)-(d), so that deviations from 1 signal deviations from the perturbative regime. 

\textit{\color{blue} Summary and experimental outlook.} We have developed a formalism which captures on equal footing the perturbative regime of fast relaxation ($e \mathbf{E}\bra{c}\mathbf{r}\ket{v}\ll\Gamma$) and the non-perturbative regime of strong light intensity ($ e \mathbf{E}\bra{c}\mathbf{r}\ket{v}\gg\Gamma$) of current rectification for interband transitions. In the perturbative regime, we recover the well-known behavior according to which shift currents approach a value that is independent of the relaxation rate $\Gamma$, while injection currents scale as $1/\Gamma$. Interestingly in the opposite non-perturbative clean limit of slow relaxation ($\Gamma \rightarrow0 $) the shift current vanishes, while the injection current approaches a finite value independent of $\Gamma$, but with a net current that scales as the square root of the radiation intensity, which can guide its identification in experiments. We have shown that this non-perturbative clean limit can be understood as optical Rabi oscillations synchronized with the incident radiation that realizes a time dependent generalized periodic Gibbs ensemble in a setting very different from its initial proposal.

Nodal Weyl semi-metals are promising platforms to realize the Rabi regime because their inter-band dipole matrix element diverges when approaching the Weyl node as $\bra{c}\mathbf{r}\ket{v} \propto 1/k$. As a consequence, they can access this non-perturbative regime above a light intensity that decreases with frequency, namely when $e v_0|\mathbf{E}| > \hbar \Gamma \omega$. For RhSi~\cite{PhysRevLett.119.206401,PhysRevLett.119.206402}, using $\hbar\Gamma^{-1}\approx10\mathrm{ps}$~\cite{Rees2020} we estimate that the non-perturbative Rabi regime will be accessed at light intensities above $4\times10^5$ W/cm$^2$ for a photon energy of $\hbar \omega \approx 0.5 \mathrm{eV}$, but this required light intensity can be decreased as $\omega^2$ at lower photon energies.

\textit{\color{blue} Acknowledgements.} We wish to thank Achilleas Lazarides, Kin Fai Mak and Andrea Cavalleri for valuable discussions and correspondence, and to Nikita Leppenen and Leonid Golub for sharing their calculations that helped us understand the precise connection to their work in Ref.\cite{Leppenen2019}. We acknowledge financial support from the Deutsche Forschungsgemeinschaft through SFB 1143 (project-id 247310070) and cluster of excellence ct.qmat (EXC 2147, project-id 39085490).

\section{Supplementary Information}
\subsection{Perturbation theory at small relaxations}\label{secA}

As discussed in Refs.\cite{PhysRevLett.123.246602,matsyshyn2020berry} the second order rectification conductivity in general can be separated into the following contributions:
\begin{multline}\label{fullcondensity matrix}
    \sigma^{\gamma\beta\alpha}_{(2)}(-\omega,\omega) =\sigma^{\gamma\beta\alpha}_{\mathrm{J}}(-\omega,\omega)+\sigma^{\gamma\beta\alpha}_{\mathrm{BCD}}(-\omega,\omega)+\\+\sigma^{\gamma\beta\alpha}_{\mathrm{I}}(-\omega,\omega)+\sigma^{\gamma\beta\alpha}_{\mathrm{S}}(-\omega,\omega),
\end{multline}
where J stands for ``Jerk", BCD for ``Berry curvature dipole", I for ``injection" and S for ``shift current" respectively. We consider band structure with small relaxations ($\forall n,m,n\neq m : \Gamma \ll \epsilon_{n}-\epsilon_m$). Each contribution is given by \cite{PhysRevLett.123.246602,matsyshyn2020berry}:
\begin{equation}\label{term1}
    \sigma^{\gamma\beta\alpha}_{\mathrm{J}}(-\omega,\omega) = \frac{e^3}{\hbar^2}\int\frac{d\mathbf{k}}{(2\pi)^3}\sum_{nm}\frac{\frac{\partial \epsilon_n}{\partial k^\gamma}\frac{\partial^2 }{\partial k^\alpha\partial k^\beta}f_n\delta_{nm}}{\omega^2+\Gamma^2},
\end{equation}    
\begin{multline}    
    \sigma^{\gamma\beta\alpha}_{\mathrm{BCD}}(-\omega,\omega) =-\frac12\frac{e^3}{\hbar^2}\frac{1}{\omega+i\Gamma}\int\frac{d\mathbf{k}}{(2\pi)^3}\times\\\times\sum_{nm}\hat{A}^{\gamma}_{mn}\hat{A}^{\alpha}_{nm}\frac{\partial}{\partial k^\beta}(f_{m}-f_n)+\\+\left(\begin{array}{c}\alpha\leftrightarrow \beta\\\omega\leftrightarrow-\omega\end{array}\right).
\end{multline}
\begin{multline}
    \sigma^{\gamma\beta\alpha}_{\mathrm{I}}(-\omega,\omega) = \frac{e^3}{\hbar^2}\int\frac{d\mathbf{k}}{(2\pi)^3}\times\\\times\sum_{nm}\frac{(f_{m}-f_n)\hat{A}^{\beta}_{nm}\hat{A}^{\alpha}_{mn}(\frac{\partial}{\partial k^\gamma}\epsilon_n-\frac{\partial}{\partial k^\gamma}\epsilon_m)}{(\omega-\epsilon_n+\epsilon_m)^2+\Gamma^2}+\\+\left(\begin{array}{c}\alpha\leftrightarrow \beta\\\omega\leftrightarrow-\omega\end{array}\right),
\end{multline}
\begin{multline}\label{termlast}
    \sigma^{\gamma\beta\alpha}_{\mathrm{S}}(-\omega,\omega) = \frac12\frac{e^3}{\hbar^2}\int\frac{d\mathbf{k}}{(2\pi)^3}\times\\\times\sum_{nm}\Bigg\{ \hat{A}^{\gamma}_{mn}\frac{\partial}{\partial k^\alpha}\frac{(f_{n}-f_m)\hat{A}^{\beta}_{nm}}{\omega-\epsilon_{n}+\epsilon_{m}+i\Gamma} 
    +\\+ i\frac{(f_{n}-f_m)\hat{A}^{\beta}_{nm}}{\omega  -\epsilon_{n}+\epsilon_{m}+i\Gamma}\sum_c \bigg[\hat{A}^\alpha_{mc}\hat{\bar{A}}^{\gamma}_{cn}-\hat{\bar{A}}^{\gamma}_{mc}\hat{A}^\alpha_{cn}\bigg]\Bigg\}+\\+\left(\begin{array}{c}\alpha\leftrightarrow \beta\\\omega\leftrightarrow-\omega\end{array}\right),
\end{multline}
where $\hat{\bar{A^\alpha}}_{nm} = \hat{A}^\alpha_{nm}(1-\delta_{nm})$ is off-diagonal Berry connection.

We split the conductivity above into resonant and non-resonant parts by separating the contributions into those that require a resonant condition that matches the light frequency with an energy difference in the limit $\Gamma\rightarrow0 $ (namely those containing delta functions enforcing a Fermi's Golden rule), from those that are non-resonant and contain the principal parts where the frequency is not forced to match an energy difference. The ``BCD" and ``Shift" conductivity have both resonant and non-resonant parts, which we label by ``R" and ``NR" additional subscripts and are given by:
\begin{multline}    
    \sigma^{\gamma\beta\alpha}_{\mathrm{BCD,R}}(-\omega,\omega) =\frac{\pi}{2}\frac{e^3}{\hbar^2}\delta(\omega)\int\frac{d\mathbf{k}}{(2\pi)^3}\times\\\times\sum_{n}\Omega_{n}^{\gamma\alpha}\frac{\partial}{\partial k^\beta}f_{n}+\left(\begin{array}{c}\alpha\leftrightarrow\beta\end{array}\right).
\end{multline}
\begin{multline}
    \sigma^{\gamma\beta\alpha}_{\mathrm{S,R}}(-\omega,\omega) = \frac\pi2\frac{e^3}{\hbar^2}\int\frac{d\mathbf{k}}{(2\pi)^3}\delta(\omega-\epsilon_{n}+\epsilon_{m})\times\\\times\sum_{nm}\Bigg\{ (f_{n}-f_m)\hat{A}^{\beta}_{nm} 
    i\frac{\partial}{\partial k^\alpha}\hat{A}^{\gamma}_{mn}+\\+ (f_{n}-f_m)\hat{A}^{\beta}_{nm}\sum_c \bigg[\hat{A}^\alpha_{mc}\hat{\bar{A}}^{\gamma}_{cn}-\hat{\bar{A}}^{\gamma}_{mc}\hat{A}^\alpha_{cn}\bigg]\Bigg\}+\\+\left(\begin{array}{c}\alpha\leftrightarrow \beta\\\omega\leftrightarrow-\omega\end{array}\right),
\end{multline}
where $\Omega^{\alpha\beta}_n = \partial A^\beta_n/\partial k^\alpha - \partial A^\alpha_n/\partial k^\beta= i[A^\alpha,A^\beta]_n$ is the Berry curvature of $n$-th band. 

Now, their non-resonant part is:
\begin{multline}    
    \sigma^{\gamma\beta\alpha}_{\mathrm{BCD,NR}}(-\omega,\omega) =\frac{i}{2}\frac{e^3}{\hbar^2}\mathrm{P.v.}\left\{\frac{1}{\omega}\right\}\int\frac{d\mathbf{k}}{(2\pi)^3}\times\\\times\sum_{n}\Omega_n^{\gamma\alpha}\frac{\partial}{\partial k^\beta}f_{n}+\left(\begin{array}{c}\alpha\leftrightarrow \beta\\\omega\leftrightarrow-\omega\end{array}\right).
\end{multline}
\begin{multline}
    \sigma^{\gamma\beta\alpha}_{\mathrm{S,NR}}(-\omega,\omega) = \frac12\frac{e^3}{\hbar^2}\int\frac{d\mathbf{k}}{(2\pi)^3}\mathrm{P.v.}\left\{\frac{1}{\omega-\epsilon_{n}+\epsilon_{m}}\right\}\times\\\times\sum_{nm}\Bigg\{ (f_{m}-f_n)\hat{A}^{\beta}_{nm}\frac{\partial}{\partial k^\alpha}\hat{A}^{\gamma}_{mn} 
    -\\- i(f_{m}-f_n)\hat{A}^{\beta}_{nm}\sum_c \bigg[\hat{A}^\alpha_{mc}\hat{\bar{A}}^{\gamma}_{cn}-\hat{\bar{A}}^{\gamma}_{mc}\hat{A}^\alpha_{cn}\bigg]\Bigg\}+\\+\left(\begin{array}{c}\alpha\leftrightarrow \beta\\\omega\leftrightarrow-\omega\end{array}\right).
\end{multline}

``Jerk" and ``injection" components can be viewed as purely resonant, and are given by the following expressions:
\begin{equation}\label{term11}
    \sigma^{\gamma\beta\alpha}_{\mathrm{J}}(-\omega,\omega) = \frac{\pi}{\Gamma}\frac{e^3 }{\hbar^2}\delta(\omega)\int\frac{d\mathbf{k}}{(2\pi)^3}\sum_{n}\frac{\partial \epsilon_n}{\partial k^\gamma}\frac{\partial^2f_n }{\partial k^\alpha\partial k^\beta},
\end{equation} 
\begin{multline}
    \sigma^{\gamma\beta\alpha}_{\mathrm{I}}(-\omega,\omega) = \frac{2\pi}{\Gamma}\frac{e^3}{\hbar^2}\int\frac{d\mathbf{k}}{(2\pi)^3}\sum_{nm}(f_{m}-f_n)\times\\\times\hat{A}^{\beta}_{nm}\hat{A}^{\alpha}_{mn}(\frac{\partial}{\partial k^\gamma}\epsilon_n-\frac{\partial}{\partial k^\gamma}\epsilon_m)\delta(\omega-\epsilon_n+\epsilon_m).
\end{multline}

For purposes of comparing with the current non-perturbative formalism, we write the injection and shift currents of two-band systems system predicted by perturbation theory for the monochromatic electric field of the following form:
\begin{equation}
        \mathbf{E}_{\text{total}}(\omega^\prime) = \mathbf{E}\delta(\omega+\omega^\prime)+\mathbf{E}^{*}\delta(\omega-\omega^\prime).
\end{equation}
Where DC current is defined as:
\begin{equation}
    j_{2,\mathrm{DC}} ^\gamma =\int_{-\infty}^{\infty}d\omega E^{\beta}(-\omega)E^{\alpha}(\omega)\sigma^{\gamma\beta\alpha}_{(2)}(-\omega,\omega),
\end{equation}
the injection and shift currents are:
\begin{multline}\label{psr}
    j_{\mathrm{S,R}}^\gamma = 2\pi\frac{e^3}{\hbar^2}\int\frac{d\mathbf{k}}{(2\pi)^3}(f_1-f_2)\delta (\omega+\epsilon_{1}-\epsilon_2) \times\\\times  \mathrm{Im}\left[E^{*\beta}_0E^{\alpha}_0A^\beta_{21}\left(\partial^\alpha A^\gamma_{12}-i\left[A^\alpha,\bar{A}^\gamma\right]_{12}\right)\right]
\end{multline}
\begin{multline}\label{psnr}
    j_{\mathrm{S,NR}}^\gamma = 2\frac{e^3}{\hbar^2}\int\frac{d\mathbf{k}}{(2\pi)^3}\frac{(\omega+\epsilon_{1}-\epsilon_2)(f_1-f_2)}{(\omega+\epsilon_{1}-\epsilon_2)^2+\Gamma^2}\times\\\times\mathrm{Re}\left[E^{*\beta}_0E^{\alpha}_0A^\beta_{21}\left(\partial^\alpha A^\gamma_{12}-i\left[A^\alpha,\bar{A}^\gamma\right]_{12}\right)\right],
\end{multline}
\begin{multline}\label{pir}
    j_{\mathrm{I}}^\gamma =\frac{e^3}{\hbar^2}\frac{2\pi}{\Gamma} \int\frac{d\mathbf{k}}{(2\pi)^3}\times\\\times(f_1-f_2)(v^\gamma_2-v^\gamma_1)|\mathbf{E}\cdot \mathbf{A}_{12}|^2\delta (\omega+\epsilon_{1}-\epsilon_2).
\end{multline}

BCD and Jerk rectification conductivities are low frequency Fermi surface terms, namely they vanish in the absence of Fermi surface at zero temperature and are associated with pole singularities at $\omega=0$, and therefore they are not described by the non-perturbative formalism of the main text that focuses on interband transitions. On the other hand, shift and injection can be non-zero for inter-band transitions. Notice that, interestingly, taken at face value, perturbation theory appears to predict that in the limit of small $\Gamma$ the shift terms remain finite while
injection diverges. 

\subsection{Floquet Formalism}\label{secB}
We use Floquet theory to determine the non-perturbative effect of the electric field and
couple the system to a bath that allows to sensibly describe steady state in the
presence of relaxation processes. The microscopic Hamiltonian has the form ($\hbar=e=1$, unless otherwise is stated):
\begin{multline}
    H= H_0(\mathbf{k}) +\frac{\partial H_0(\mathbf{k})}{\partial k^\alpha} A_0^\alpha(t)+\frac{1}{2}\frac{\partial^2 H_0(\mathbf{k})}{\partial k^\alpha \partial k^\beta}A_0^\alpha(t)A_0^\beta(t)+\\+\frac{1}{3!}\frac{\partial^3 H_0(\mathbf{k})}{\partial k^\alpha \partial k^\beta \partial k^\gamma}A_0^\alpha(t)A_0^\beta(t)A_0^\gamma(t)+\mathcal{O}(E^4).
\end{multline}

Expansion to 3rd order in vector potentials is necessary to compute electric currents correctly to
order $E_0^2$, since the current is the expectation value of the velocity operator:
\begin{multline}
    v^\lambda = \frac{\partial H_0}{\partial k^\lambda} +\frac{\partial^2 H_0(\mathbf{k})}{\partial k^\alpha \partial k^\lambda} A_0^\alpha(t)+\\+\frac{1}{2}\frac{\partial^3 H_0(\mathbf{k})}{\partial k^\alpha \partial k^\beta \partial k^\lambda}A_0^\alpha(t)A_0^\beta(t)+\mathcal{O}(E^3).
\end{multline}

We assume that the electric field is a general periodic function with frequency $\omega$:
\begin{gather}
     \mathbf{A}_0(t) = i\frac{\mathbf{E}}{\omega}e^{i\omega t}-i\frac{\mathbf{E}^{*}}{\omega}e^{-i\omega t},\\
    \mathbf{E}_{\text{total}}(t) =-\frac{\partial \mathbf{A}(t)}{\partial t}= \mathbf{E}e^{i\omega t}+\mathbf{E}^{*}e^{-i\omega t}.
\end{gather}

Since the Hamiltonian is periodic ($\omega = 2\pi/T$), we apply discrete Fourier transform and now the Hamiltonian in Floquet picture has the following components:
\begin{gather}
    H(t) = \sum_{n=-\infty}^{\infty}H^f_ne^{in\omega t},\quad H^f_n = \frac{1}{T}\int_0^{T} H e^{-in\omega t} dt,\\
    H^f_0 = H_0(\mathbf{k}) +e^2\mathrm{Re}\left[\frac{E^\alpha E^{*\beta}}{\omega^2}\right]\frac{\partial^2 H_0(\mathbf{k})}{\partial k^\alpha \partial k^\beta},
\end{gather}
\begin{equation}
    H_1^f = i\frac{E^\alpha}{\omega}\frac{\partial H_0(\mathbf{k})}{\partial k^\alpha} +i\frac{\partial^3 H_0(\mathbf{k})}{\partial k^\alpha \partial k^\beta\partial k^\lambda}\frac{E_0^\lambda}{\omega}\mathrm{Re}\left[\frac{E^\alpha E^{*\beta}}{\omega^2}\right]
\end{equation}
\begin{multline}
    H_{-1}^f = -i\frac{E^{*\alpha}}{\omega}\frac{\partial H_0(\mathbf{k})}{\partial k^\alpha} -\\-i\frac{\partial^3 H_0(\mathbf{k})}{\partial k^\alpha \partial k^\beta\partial k^\lambda}\frac{E^{*\lambda}}{\omega}\mathrm{Re}\left[\frac{E^\alpha E^{*\beta}}{\omega^2}\right],
\end{multline}
\begin{gather}    
    H_2^f = -\frac{E^\alpha E^\beta}{\omega^2}\frac12\frac{\partial^2 H_0(\mathbf{k})}{\partial k^\alpha \partial k^\beta}, \\ H_{-2}^f = -\frac{E^{*\alpha} E^{*\beta}}{\omega^2}\frac12\frac{\partial^2 H_0(\mathbf{k})}{\partial k^\alpha \partial k^\beta},\\
    H_3^f = -i\frac{E^\alpha E^\beta E^\lambda}{\omega^3}\frac{1}{3!}\frac{\partial^3 H_0(\mathbf{k})}{\partial k^\alpha \partial k^\beta\partial k^\lambda}, \\ H_{-3}^f = i\frac{E^{*\alpha} E^{*\beta}E^{*\lambda}}{\omega^3}\frac{1}{3!}\frac{\partial^3 H_0(\mathbf{k})}{\partial k^\alpha \partial k^\beta\partial k^\lambda}.
\end{gather}

Thus the Hamiltonian structure in Floquet picture is:
\begin{multline}\label{steptruncation}
    \hat{H}_{\mathbf{F}}=\\\left(\begin{array}{cccccccc}\ddots&\vdots&\vdots&\vdots&\vdots&\vdots\\
    \cdots&H_0+2\omega&H_1&H_2&H_3&0&\cdots\\\cline{3-4}
    \cdots&H_{-1}&\multicolumn{1}{|c}{H_0+\omega} & \multicolumn{1}{c|}{H_1} &H_2&H_3&\cdots\\\cdots&H_{-2}&
    \multicolumn{1}{|c}{H_{-1}}&\multicolumn{1}{c|}{H_0}&H_1&H_2&\cdots\\\cline{3-4}
    \cdots&H_{-3}&H_{-2}&
    H_{-1}&H_0-\omega&H_1&\cdots\\\cdots&0&H_{-3}&H_{-2}&
    H_{-1}&H_0-2\omega&\cdots\\&\vdots&\vdots&\vdots&\vdots&\vdots&\ddots\end{array}\right)
\end{multline}

Additionally, one can show that in Floquet representation the velocity operator has the following form:
\begin{gather}
    v^\lambda_0 = \frac{\partial H_0}{\partial k^\lambda} +\frac{\partial^3 H_0(\mathbf{k})}{\partial k^\alpha \partial k^\beta \partial k^\lambda}\mathrm{Re}\left[\frac{E^\alpha E^{*\beta}}{\omega^2}\right],\\
    v_1^\lambda = i\frac{E^\alpha}{\omega}\frac{\partial^2 H_0(\mathbf{k})}{\partial k^\alpha\partial k^\lambda},\quad v_{-1}^\lambda = -i\frac{E^{*\alpha}}{\omega}\frac{\partial^2 H_0(\mathbf{k})}{\partial k^\alpha\partial k^\lambda}\\ v_2^\lambda = -\frac{1}{2} \frac{\partial^3 H_0(\mathbf{k})}{\partial k^\alpha \partial k^\beta \partial k^\lambda}E^\alpha E^\beta,\\
    v_{-2}^\lambda = -\frac{1}{2} \frac{\partial^3 H_0(\mathbf{k})}{\partial k^\alpha \partial k^\beta \partial k^\lambda}E^{*\alpha}E^{*\beta}.
\end{gather}

We follow the approach of Ref.\cite{Morimotoe1501524} and take a simplified 2 band
model. We focus on inter-band resonant processes between the conduction and valence bands. The matrix elements highlighted by the rectangular box in Eq.(\ref{steptruncation}) are the ones that we keep within the ``rotating wave" two band truncation, and ignore contributions of order $\mathbf{E}^3$ and higher. Therefore,  the Floquet Hamiltonian and velocity operator can be truncated to an effective 2 band model, which reads as:
\begin{gather}
    H^\mathbf{F}_{\mathrm{T}} = 
 \left(\begin{array}{cc}\epsilon_1+\omega&i(\mathbf{A}\cdot \mathbf{v}_{12})\\ 
 -i(\mathbf{A}^*\cdot \mathbf{v}_{21})&\epsilon_2\end{array}\right) = h_0+\mathbf{h}\cdot\mathbf{\sigma},\\\label{HammFloq}
     v^{\mathbf{F},\alpha}_T = \left(\begin{array}{cc}
         \partial^\alpha E_1& i\mathbf{A}\cdot \left(\frac{\partial^2 H_0(\mathbf{k})}{\partial k^\alpha\partial\mathbf{k}}\right)_{12} \\
         -i\mathbf{A}^{*}\cdot \left(\frac{\partial H_0(\mathbf{k})}{\partial k^\alpha\partial \mathbf{k}}\right)_{21}&\partial ^\alpha E_2
    \end{array} \right)+v^{\alpha}_{E^2},\\
    (v^{\lambda}_{E^2})_{i,j} = \left(\frac{\partial^3 H_0(\mathbf{k})}{\partial k^\alpha \partial k^\beta \partial k^\lambda}\right)_{ij}\mathrm{Re}\left[\frac{E^\alpha E^{*\beta}}{\omega^2}\right],\\
    h_0 = \frac12(\epsilon_1+\epsilon_2+\omega),\qquad h_x = \mathrm{Re}[i(\mathbf{A}^*\cdot \mathbf{v}_{21})],\\ h_y = \mathrm{Im}[i(\mathbf{A}^*\cdot \mathbf{v}_{21})],\qquad  h_z =\frac12(\epsilon_1-\epsilon_2+\omega),\\
    \epsilon_1 = E_{1}+\mathrm{Re}\left[\frac{E^\alpha E^{*\beta}}{\omega^2}\right]\left(\frac{\partial^2 H_0}{\partial k^\alpha \partial k^\beta}\right)_{11}, \\ \epsilon_2 = E_{2}+\mathrm{Re}\left[\frac{E^\alpha E^{*\beta}}{\omega^2}\right]\left(\frac{\partial^2 H_0}{\partial k^\alpha \partial k^\beta}\right)_{22},\label{HammFloqf} \\
    \mathbf{v}_{12} = \left(\frac{\partial H_0(\mathbf{k})}{\partial \mathbf{k}}\right)_{12},\qquad  \mathbf{v}_{21} = \left(\frac{\partial H_0(\mathbf{k})}{\partial \mathbf{k}}\right)_{21},
\end{gather}
where 1 stands for ``valence", 2 for ``conduction ", ($i,j$) denotes the band index, $E_{1,2}$ are unperturbed band energies, $\epsilon_{1,2}$ are dressed effective band energies and $\mathbf{A}=\mathbf{E}/\omega$. The above formalism is a slight generalization of that in Ref.\cite{Morimotoe1501524}, that adds some dressing of the band energies by the drive, although this ingredient is not crucial for the key quantitative predictions made in the main text.

Within our current notation, the following is the convention to convert a matrix from Floquet picture into the usual Schrödinger picture:
\begin{equation}
    \hat{O}_\mathbf{F} = \left(\begin{array}{cc}
        O_{22} & O_{12}  \\
        O_{21} & O_{11}
    \end{array}\right),\qquad
    \hat{O}(t) = \left(\begin{array}{cc}
        O_{11} & O_{21}e^{-i\omega t}  \\
        O_{12}e^{i\omega t} & O_{22}
    \end{array}\right).
\end{equation}

And for the states (eigenvectors of the Hamiltonian from Eq.(\ref{HammFloq})), the Floquet representation and the Schrödinger picture are related as:
\begin{equation}
    \psi^j_{\mathbf{F}}=\left(\begin{array}{c}u_1^{j}\\u_2^{j}\end{array}\right)\quad\longrightarrow \quad\psi_{j}(t) = e^{-i\epsilon_{j}t}\left(\begin{array}{c}u_2^{j}\\u_1^{j}e^{i\omega t}\end{array}\right).
\end{equation}

\subsection{General Keldysh Formalism}\label{secC}
To model relaxation we couple our system to a simple Fermion bath at temperature T and chemical potential $\epsilon_F$. The bath couples uniformly to each site of the lattice sites where the fermions hop, as depicted in Fig.1(b). The partition function and Lagrangian of the system are given by:
\begin{gather}
    Z = \frac{\mathrm{Tr}[\rho U_c]}{\mathrm{Tr}[\rho]}= \int D[a,a^\dagger,c,c^\dagger]\rho_0e^{i\mathcal{S}},\\
    \mathcal{S} = \int_{\mathcal{C}}dt \left\{\mathcal{L}_{\mathrm{TB}}+\mathcal{L}_{\mathrm{mix}}+\mathcal{L}_{\mathrm{bath}}\right\},\\
    \mathcal{L}_{\mathrm{TB}}=\sum_{i\neq j}c_i^\dagger(i\partial_t - H^{ij})c_j,\\
    \mathcal{L}_{\mathrm{mix}}=\sum_{k,i}V^k_{\mathrm{mix}}\left[a_k^\dagger c_i + \mathrm{h.c.}\right],\\
    \mathcal{L}_{\mathrm{\mathrm{bath}}}=\sum_k a_k^\dagger (i\partial_t - \epsilon_k^{\mathrm{bath}})a_k,\\ \langle a^\dagger_k a_k\rangle = n(k)=f (\epsilon_k^{\mathrm{bath}})= [1+e^{\beta(\epsilon_k^{\mathrm{bath}}-\epsilon_F)}]^{-1},
\end{gather}
where $\mathcal{L}_{\mathrm{TB}}$ describes the tight binding model of the system without the bath, $\mathcal{L}_{\mathrm{\mathrm{mix}}}$ the tunneling from the system sites to the states in the bath, and $\mathcal{L}_{\mathrm{\mathrm{bath}}}$ describes the bath. We will take $\rho_0$ to be the initial density matrix of the whole system plus bath, ($a_k,a_k^\dagger$) are creation and annihilation operators of the bath states labeled by $k$, ($c_i,c_i^\dagger$) is creation and annihilation operator of the fermion on the site ``$i$'', $H^{ij}$ is the tight-binding Hamiltonian of the system, $\epsilon_k^{\mathrm{bath}}$ is the energy of the k-th state of the bath, $\beta = 1/(k_BT)$, $V^k_{\mathrm{mix}}$ is the coupling of the system site (all sites of the physical system are coupled to identical and independent baths, as depicted in Fig.1(b)) to the k-th state of the bath and $f(\omega)$ is the Fermi-Dirac distribution. $\mathcal{C}$ is the closed contour for the time integration of the Keldysh approach \cite{KelFloq1}. 

We can split the closed contour of time integration into two parts:
\begin{equation}
    \int_{\mathcal{C}} = \int_{-\infty}^{\infty}dt\left[(\cdots)_+ - (\cdots)_-\right],
\end{equation}
the sub-script ``+" or ``-" denote the forward and backward parts of the contour and plays the role of an additional effective internal degree of freedom~\cite{kamenev_2011}. From this one can write the action describing the system-bath coupling as follows: 
\begin{multline}
    \mathcal{S}_{\mathrm{mix}}=\\=\int_{-\infty}^{\infty} dt\sum_{k,i}V^k_{\mathrm{mix}}\Big[(a_{k,+}^\dagger c_{i,+} + \mathrm{h.c.}) - (a_{k,-}^\dagger c_{i,-} + \mathrm{h.c.}) \Big]=\\= \int_{-\infty}^{\infty}dt\sum_{k,i}V^k_{\mathrm{mix}}\left[\left(\begin{array}{cc}
         a_{k,+}^\dagger&a_{k,-}^\dagger   
    \end{array}\right)\left(\begin{array}{cc}
         1&0  \\
         0&-1 
    \end{array}\right)\left(\begin{array}{cc}
          c_{i,+} \\c_{i,-}
    \end{array}\right)+\mathrm{h.c}\right]
\end{multline}

Now we apply the standard Keldysh fermionic rotation:
\begin{gather}
    \left(\begin{array}{cc}
           c_{i,+}\\c_{i,-}
    \end{array}\right)  = \frac{1}{\sqrt{2}}\left(\begin{array}{cc}
         1&1  \\
         1&-1 
    \end{array}\right) \left(\begin{array}{cc}
           c_{i,1}\\c_{i,2}
    \end{array}\right),\\
        \left(\begin{array}{cc}
           a_{k,+}^\dagger&a_{k,-}^\dagger
    \end{array}\right)  = \left(\begin{array}{cc}
           a_{k,1}^\dagger&a_{k,2}^\dagger
    \end{array}\right)\frac{1}{\sqrt{2}}\left(\begin{array}{cc}
         1&-1  \\
         1&1 
    \end{array}\right).
\end{gather}
After Fourier transforming the above we get:
\begin{equation}
    \mathcal{S}_{\mathrm{mix}}=\int_{-\infty}^{\infty}d\omega\sum_{k,i}V^k_{\mathrm{mix}}\left[\left(\begin{array}{cc}
         a_{k,1}^\dagger&a_{k,2}^\dagger   
    \end{array}\right)\left(\begin{array}{cc}
          c_{i,1} \\c_{i,2}
    \end{array}\right)+\mathrm{h.c}\right].
\end{equation}

Applying the same procedure to the rest of the action brings us to the following result:
\begin{multline}
    \mathcal{S}_{\mathrm{\mathrm{bath}}}= \int_{-\infty}^{\infty} d\omega\sum_k  \left(\begin{array}{cc}
         a_{k,1}^\dagger&a_{k,2}^\dagger   
    \end{array}\right)\times\\\times \left(\begin{array}{cc}
         G^\mathbf{R}(k,\omega)&G^{\mathbf{K}}(k,\omega) \\
         0&G^\mathbf{A}(k,\omega) 
    \end{array}\right)^{-1}\left(\begin{array}{cc}
          a_{k,1} \\a_{k,2}
    \end{array}\right),
\end{multline}
\begin{multline}
    \mathcal{S}_{\mathrm{TB}} = \sum_{ij}\int_{-\infty}^{\infty} d\omega \left(\begin{array}{cc}
         c_{i,1}^\dagger&c_{i,2}^\dagger   
    \end{array}\right)\times\\\times \left(\begin{array}{cc}
         \omega- H^{ij}(\omega)+i\varepsilon&2i\varepsilon f^\mathbf{e}(\omega)\\
         0&\omega - H^{ij}(\omega)-i\varepsilon 
    \end{array}\right)\left(\begin{array}{cc}
          c_{j,1} \\c_{j,2}
    \end{array}\right),
\end{multline}
where $G^{\mathbf{A}/\mathbf{R}}$ are advanced and retarded Green functions respectively and $G^{\mathbf{K}}$ stands for the Keldysh Green function. They are given by:
\begin{gather}
    G^{\mathbf{R}}(k,\omega) = (\omega - \epsilon^{\mathrm{bath}}_k + i0)^{-1},\\
    G^{\mathbf{A}}(k,\omega) = (\omega - \epsilon^{\mathrm{bath}}_k - i0)^{-1},\\
    G^{\mathbf{K}}(k,\omega) =-2\pi i(1 - 2 f(\omega))\delta(\omega - \epsilon^{\mathrm{bath}}_k), 
\end{gather}
where $f (\omega)= (1+e^{\beta(\omega-\epsilon_F)})^{-1}$.

After integrating out the fermionic bath, we obtain the following effective action for the system:
\begin{equation}
    \mathcal{S}_{\mathrm{eff}} = \mathcal{S}_{\mathrm{TB}}+\mathcal{S}_{\mathrm{int}}.
\end{equation}
\begin{multline}
    \mathcal{S}_{\mathrm{\mathrm{int}}}=-\int_{-\infty}^{\infty} d\omega\sum_{k,i}|V^k_{\mathrm{mix}}|^2 \times\\\times\left(\begin{array}{cc}
         c_{i,1}^\dagger&c_{i,2}^\dagger   
    \end{array}\right)\left(\begin{array}{cc}
         G^\mathbf{R}(k,\omega)&G^{\mathbf{K}}(k,\omega) \\
         0&G^\mathbf{A}(k,\omega) 
    \end{array}\right)\left(\begin{array}{cc}
          c_{i,1} \\c_{i,2}
    \end{array}\right).
\end{multline}

In order to obtain the equal time lesser GF:
\begin{equation}
    G^< = \frac{G^{\mathbf{K}}+G^{\mathbf{A}}-G^{\mathbf{R}}}{2},
\end{equation}
we can focus on the imaginary parts of the Green functions:
\begin{gather}
    G^{\mathbf{R}}(k,\omega) =-i\pi\delta(\omega - \epsilon^{\mathrm{bath}}_k),\\
    G^{\mathbf{A}}(k,\omega) = i\pi\delta(\omega - \epsilon^{\mathrm{bath}}_k)_k),\\
    G^{\mathbf{K}}(k,\omega) =-2\pi i(1 - 2 f(\omega))\delta(\omega - \epsilon^{\mathrm{bath}}_k), 
\end{gather}
which leads us to the following form of effective interaction:
\begin{multline}
    \mathcal{S}_{\mathrm{\mathrm{int}}}=\int_{-\infty}^{\infty} d\omega\sum_{i}\times\\\times\left(\begin{array}{cc}
         c_{i,1}^\dagger&c_{i,2}^\dagger   
    \end{array}\right)\left(\begin{array}{cc}
         i\Gamma(\omega)/2&i\Gamma(\omega)(1-2f(\omega)) \\
         0&-i\Gamma(\omega)/2 
    \end{array}\right)\left(\begin{array}{cc}
          c_{i,1} \\c_{i,2}
    \end{array}\right),
\end{multline}
where $\Gamma(\omega)$ is an effective imaginary self-energy or relaxation rate, that captures the relaxation processes and it is explicitly given by:
\begin{equation}\label{GammaBath}
    \Gamma(\omega) = 2\pi\sum_k |V^k_{\mathrm{mix}}|^2 \delta(\omega - \epsilon^{\mathrm{bath}}_k).
\end{equation}

\subsection{Keldysh Floquet Formalism}
In the case of a periodically driven system the action satisfies:
\begin{gather}
    \mathcal{S}_{\mathrm{TB}} = \sum_{ij}\int dt c_i^\dagger(t) (i\partial_t - H^{ij}(t)\pm i0)c_j(t),\\
    H^{ij}(t) = H^{ij}(t+T).
\end{gather}
Therefore we can expand it with a the discrete Fourier transform:
\begin{gather}
    H^{ij}(t) = \sum_{m=-\infty}^{\infty}H^{ij}_{m}e^{-im\omega t},\\
    H^{ij}_{m} = \frac{1}{T}\int_0^T dt H^{ij}(t) e^{im\omega t},
\end{gather}
and arrive to the following action:
\begin{multline}
    \mathcal{S}_{\mathrm{TB}} = \sum_{ij}\sum_{m^\prime=-\infty}^{\infty}\int_{-\infty}^\infty d\omega \times\\\times c_i^\dagger(\omega) (\omega\delta_{m^\prime,0} - H^{ij}_{m^\prime}\pm i0)c_j(\omega+m^\prime\omega).
\end{multline}

Additionally, after splitting the frequency integration into segments:
\begin{equation}
    \int_{-\infty}^{\infty}d\omega^\prime f(\omega^\prime) = \sum_{n= -\infty}^{\infty}\int_{-\omega/2}^{\omega/2}d\omega^\prime f(\omega^\prime +n \omega),
\end{equation}
where $\omega = 2\pi/T$, and relabeling the summation index ($n+m^\prime = m$), one obtains the result:
\begin{multline}
    \mathcal{S}_{\mathrm{TB}} = \sum_{ij}\sum_{n,m=-\infty}^{\infty}\int_{-\omega/2}^{\omega/2}d\omega^\prime c_i^\dagger(\omega^\prime +n\omega)\times\\\times  ((\omega^\prime+n\omega)\delta_{n,m} - H^{ij}_{m-n}\pm i0)c_j(\omega^\prime+m\omega).
\end{multline}

And analogous procedure can be performed with the tight binding part of the action, yielding:
\begin{multline}
        \mathcal{S}_{\mathrm{TB}} = \sum_{ij}\sum_{n,m=-\infty}^{\infty}\int_{-\omega/2}^{\omega/2}d\omega^\prime\left(\begin{array}{cc}
         c_{i,1,n}^\dagger&c_{i,2,n}^\dagger   
    \end{array}\right)\times\\(\delta_{nm}\omega^\prime\hat{\mathds{1}}-\hat{H}^{\mathbf{F},ij}_{mn})\left(\begin{array}{cc}
          c_{i,1,m} \\c_{i,2,m}
    \end{array}\right),
\end{multline}
\begin{equation}
    \hat{H}^{\mathbf{F},ij}_{mn} =\left(\begin{array}{cc}
           H^{ij}_{m-n}-n\omega\delta_{nm}&2i\varepsilon f^\mathbf{e}(\omega^\prime+n\omega)\\
         0& H^{ij}_{m-n}-n\omega\delta_{nm}
    \end{array}\right),
\end{equation}

\begin{multline}
    \mathcal{S}_{\mathrm{int}}=\sum_{i}\sum_{n=-\infty}^{\infty}\int_{-\omega/2}^{\omega/2}d\omega\sum_{i}\left(\begin{array}{cc}
         c_{i,1,n}^\dagger&c_{i,2,n}^\dagger   
    \end{array}\right)\times\\\times\left(\begin{array}{cc}
         i\Gamma(\omega^\prime+n\omega)/2&i\Gamma(\omega^\prime+n\omega)(1-2f(\omega^\prime+n\omega)) \\
         0&-i\Gamma(\omega^\prime +n\omega)/2 
    \end{array}\right)\times\\\times\left(\begin{array}{cc}
          c_{i,1,n} \\c_{i,2,n}
    \end{array}\right),
\end{multline}
where $c_{i,a,n} = c_{i,a}(\omega + n\omega)$. 

Now the system's full Green function, after integrating out the bath, is given by:
\begin{multline}
    \mathbf{G}^{-1} = \left(\begin{array}{cc}
          G^{\mathbf{R}}&G^{\mathbf{K}} \\0&G^{\mathbf{A}}
    \end{array}\right)^{-1}_{ij,nm} =\\ \left(\begin{array}{cc}
         (\omega^\prime+n\omega)\delta^{ij}_{nm} - H^{ij}_{m-n}&2i\varepsilon f^\mathbf{e}(\omega)\\
         0&(\omega^\prime+n\omega)\delta^{ij}_{nm} - H^{ij}_{m-n}
    \end{array}\right) +\\\delta^{ij}_{nm}\left(\begin{array}{cc}
         i\Gamma(\omega^\prime+n\omega)/2&i\Gamma(\omega^\prime+n\omega)(1-2f(\omega^\prime+n\omega)) \\
         0&-i\Gamma(\omega^\prime+n\omega)/2 
    \end{array}\right),
\end{multline}
where $\delta^{ij}_{nm} = \delta_{ij}\delta_{nm}$ are Kronecker deltas.

The lesser Green Function $G^<$ is the one we need in order to obtain the density matrix of the system, which can related to advanced $G^\mathbf{A}$ and retarded $G^\mathbf{R}$ Greens functions, as follows:
\begin{gather}
    \mathbf{G}^{-1} = G^{-1}_0 - \Sigma,\\
    \Sigma = \left(\begin{array}{cc}
         \Sigma^R&\Sigma^K  \\
         0& \Sigma^A
    \end{array}\right),\qquad\Sigma^< = \frac{\Sigma^{\mathbf{K}}+\Sigma^{\mathbf{A}}-\Sigma^{\mathbf{R}}}{2},\\
    G^{<} = G^\mathbf{R} \Sigma^{<} G^\mathbf{A}.
\end{gather}

We have:
\begin{equation}\label{sigmasles}
    \Sigma^{<,ij}_{nm} (\omega^\prime)= \delta_{ij}\delta_{nm} f(\omega^\prime+n\omega). 
\end{equation}
Therefore the technical task is reduced to finding the advanced and retarded green functions of the system. 

\subsection{Truncated Green Functions}\label{secE}
For the special choice of bath in Fig.1(b) in which each site couples to an identical bath, the self energy from Eq.(\ref{sigmasles}) is independent of the system site indices ``$ij$". To obtain the same truncation, in the spirit of the rotating wave approximation, as we had before for Eq.(\ref{HammFloq}), here we need to restrict to the indices ($n = 0,n = -1$). Thus effectively, after truncation we have:
\begin{gather}
    \Sigma^<(\omega^\prime) = i\Gamma(\omega^\prime)\left(\begin{array}{cc}
         f_1(\omega^\prime-\omega)&0  \\
         0& f_2(\omega^\prime)
    \end{array}\right),\\\Sigma^<_v(\omega^\prime) = i\Gamma(\omega^\prime)\left(\begin{array}{cc}
         f_1(\omega^\prime-\omega)&0  \\
         0& 0
    \end{array}\right),\\ \Sigma^<_c(\omega^\prime) = i\Gamma(\omega^\prime)\left(\begin{array}{cc}
         0&0  \\
         0& f_2(\omega^\prime)
    \end{array}\right),
\end{gather}
and
\begin{gather}
    G^{\mathbf{R}^{-1}}_{\mathrm{S}} = (\omega^\prime+i\Gamma(\omega^\prime)/2-h_0)\mathbb{I}-\mathbf{h}\cdot\mathbf{\sigma},\\
    G^{\mathbf{A}^{-1}}_{\mathrm{S}} = (\omega^\prime-i\Gamma(\omega^\prime)/2-h_0)\mathbb{I}-\mathbf{h}\cdot\mathbf{\sigma},
\end{gather}
all the notations are defined in Eq.(\ref{HammFloq}-\ref{HammFloqf}). 

After inverting we have:
\begin{gather}
    G^{\mathbf{R}}_{\mathrm{T}} (\omega^\prime)=\frac{ (\omega^\prime+i\Gamma(\omega^\prime)/2-h_0)\mathbb{I}+\mathbf{h}\cdot\mathbf{\sigma}}{(\omega+i\Gamma(\omega^\prime)/2-h_0)^2-h^2},\\
    G^{\mathbf{A}}_{\mathrm{T}}(\omega^\prime) =\frac{ (\omega^\prime-i\Gamma(\omega^\prime)/2-h_0)\mathbb{I}+\mathbf{h}\cdot\mathbf{\sigma}}{(\omega^\prime-i\Gamma(\omega^\prime)/2-h_0)^2-h^2}
\end{gather}

The lesser GFs associated with ``valence" and ``conduction" bands are given by: 
\begin{equation}
    G^<_v = G^{\mathbf{R}}\Sigma^<_v G^{\mathbf{A}},\qquad G^<_c = G^{\mathbf{R}}_{\mathrm{S}}\Sigma^<_c G^{\mathbf{A}}_{\mathrm{S}}.
\end{equation}

Or explicitly:
\begin{multline}
    G^<_v(\omega^\prime) =f_1(\omega^\prime-\omega)\Gamma(\omega^\prime)\times\\\frac{ \left(\begin{array}{cc}i(\omega^\prime-h_0+h_z)^2+i\Gamma^2/4 & h_-(ih_z-\Gamma/2+i(\omega^\prime-h_0))\\h_+(ih_z+\Gamma/2+i(\omega^\prime-h_0))&i(h_x^2+h_y^2)\end{array}\right)}{((\omega^\prime+i\Gamma/2-h_0)^2-h^2)((\omega^\prime-i\Gamma/2-h_0)^2-h^2)},
\end{multline}

\begin{multline}
    G^<_c(\omega^\prime) = f_2(\omega^\prime)\Gamma(\omega^\prime)\times\\\frac{ \left(\begin{array}{cc}i(h_x^2+h_y^2) & h_-(\Gamma/2-ih_z+i(\omega^\prime-h_0))\\h_+(-\Gamma/2-ih_z+i(\omega^\prime-h_0))&i(\omega^\prime-h_0-h_z)^2+i\Gamma^2/4\end{array}\right)}{((\omega^\prime+i\Gamma/2-h_0)^2-h^2)((\omega^\prime-i\Gamma/2-h_0)^2-h^2)}.
\end{multline}

In general, the time average density matrix is given by:
\begin{equation}
    \hat{\rho}_{\mathbf{DC}} = -i\int_{-\infty}^{\infty}\frac{d\omega^\prime}{2\pi}\left(G^<_c(\omega^\prime)+G^<_v(\omega^\prime)\right).
\end{equation}

To proceed analytically, we assume a bath with a broad and flat density of states over the system's energy states, which allows us to neglect the frequency dependence of relaxation rate in  Eq.(\ref{GammaBath}), and therefore we take $\Gamma$ as a constant. Additionally, we assume that the dressed band energies are above or below Fermi energy, which allows us to take the occupation numbers as frequency independent (this is rigorously justified only in insulators, but in the case of metals, it can be viewed as a reasonable approximation that will captures the essence of the Pauli blocking effect of optical transitions). After frequency integration we have:
\begin{equation}
    G^<_v =f_1\frac{ \left(\begin{array}{cc}ih^2+ih_z^2+i\Gamma^2/2 & h_-(ih_z-\Gamma/2)\\h_+(ih_z+\Gamma/2)&i(h_x^2+h_y^2)\end{array}\right)}{2(h^2+\frac{\Gamma^2}{4})}
\end{equation}

\begin{equation}
    G^<_c =f_2\frac{ \left(\begin{array}{cc}i(h_x^2+h_y^2) & -h_-(ih_z-\Gamma/2)\\-h_+(ih_z+\Gamma/2)&ih^2+ih_z^2+i\Gamma^2/2\end{array}\right)}{2(h^2+\frac{\Gamma^2}{4})}
\end{equation}

The DC current is:
\begin{equation}
    \mathbf{J} = -i\mathbf{Tr}(\hat{\mathbf{v}}_{\mathbf{F}}G^<),\qquad G^<=G^<_c+G^<_v,
\end{equation}
where the velocity operator is:
\begin{gather}
    v^{\mathbf{F},\alpha}_T = \left(\begin{array}{cc}
         \partial^\alpha E_1& -i\mathbf{A}\cdot \left(\frac{\partial^2 H_0(\mathbf{k})}{\partial k^\alpha\partial\mathbf{k}}\right)_{12} \\
         i\mathbf{A}^{*}\cdot \left(\frac{\partial H_0(\mathbf{k})}{\partial k^\alpha\partial \mathbf{k}}\right)_{21}&\partial ^\alpha E_2
    \end{array} \right)+v^{\alpha}_{E^2},\\
    (v^{\alpha}_{E^2})_{ij} = \left(\frac{\partial^3 H_0(\mathbf{k})}{\partial k^\alpha \partial k^\beta \partial k^\lambda}\right)_{ij}\mathrm{Re}\left[\frac{E^\alpha E^{*\beta}}{\omega^2}\right].
\end{gather}

Thus the explicit result is given by:
\begin{multline}
    J^\alpha = -i\mathbf{Tr}(\hat{v}^\alpha_{\mathbf{F}}G^<) =\\= -i\mathbf{Tr}((v^\alpha_0+\mathbf{v}^\alpha\cdot\mathbf{\sigma})G^<)=\sum_{\mathbf{k}}\frac{(f_1-f_2)}{h^2+\frac{\Gamma^2}{4}}\times\\\times\left[\underbrace{\frac{\Gamma}{2}(h_yv^\alpha_x-v^\alpha_yh_x)}_{J_1}+\underbrace{h_z(v^\alpha_xh_x+v^\alpha_yh_y)}_{J_2}+\underbrace{v^\alpha_z\left(h_z^2+\frac{\Gamma^2}{4}\right)}_{J_3}\right]+\\+v_0(f_1+f_2),
\end{multline}
where $\sum_{\mathbf{k}} = \int d^dk/(2\pi)^d$ and the lower ($x,y,z$) index denotes the decomposition of the matrix in $\sigma$-basis. The last term in the expression above is effectively an average of the velocity operator in the quasi-equilibrium state with dressed energies, and can be seen to identically vanish.

Now, we split the current into on three contributions:
\begin{gather}
    J_1^\alpha = \sum_{\mathbf{k}}\frac{\frac{\Gamma}{2}(f_{1}-f_2)}{h^2+\Gamma^2/4}(h_yv^\alpha_x-v^\alpha_yh_x),\label{kfcur1}\\
    J_2^\alpha =\sum_{\mathbf{k}}\frac{h_z(f_{1}-f_2)}{h^2+\Gamma^2/4}(h_xv_x^\alpha+h_yv_y^\alpha),\\
    J_3^\alpha = -\sum_{\mathbf{k}}(f_1-f_2)v^\alpha_z\frac{h_x^2+h_y^2}{h^2+\frac{\Gamma^2}{4}},\label{kfcur3}
\end{gather}
where $v_{2}= v_0-v_z, v_1 = v_0+v_z$. Note that $h_{x,y} \sim E_0$, which means that in order to compute the currents ($J_1, J_2, J_3$) with accuracy $\mathcal{O}(E_0^2)$, we can safely assume that ${\hat{v}}^\alpha_{\mathbf{F}} = \partial \hat{H}/\partial {k}^\alpha$ and ignore the $v^{\alpha}_{E^2}$ contribution. Finally, the velocity operator (with restored units) can be approximated as:
\begin{multline}\label{velocityfloq}
    v^{\mathbf{F},\alpha}_T =  \left(\begin{array}{cc}
         \partial^\alpha E_1& i\frac{e}{\hbar}\mathbf{A}\cdot \left(\frac{\partial^2 H_0(\mathbf{k})}{\partial k^\alpha\partial\mathbf{k}}\right)_{12} \\
         -i\frac{e}{\hbar}\mathbf{A}^{*}\cdot \left(\frac{\partial H_0(\mathbf{k})}{\partial k^\alpha\partial \mathbf{k}}\right)_{21}&\partial ^\alpha E_2
    \end{array} \right)+\\\\
    e^2\mathrm{Re}\left[\frac{E^\alpha E^{*\beta}}{\hbar^2\omega^2}\right]\left(\begin{array}{cc}
         \left(\frac{\partial^3 H_0(\mathbf{k})}{\partial k^\alpha\partial k^\beta\partial k^\lambda}\right)_{11}& 0 \\
         0& \left(\frac{\partial^3 H_0(\mathbf{k})}{\partial k^\alpha\partial k^\beta\partial k^\lambda}\right)_{22}
    \end{array}\right).
\end{multline}

Or in the Schrödinger picture:
\begin{multline}\label{velocitytime}
    v^{\alpha}(t) = \\\left(\begin{array}{cc}
         \partial^\alpha E_2& -i\frac{e}{\hbar}\mathbf{A}^*\cdot \left(\frac{\partial^2 H_0(\mathbf{k})}{\partial k^\alpha\partial\mathbf{k}}\right)_{21}e^{-i\omega t} \\
         i\frac{e}{\hbar}\mathbf{A}\cdot \left(\frac{\partial^2 H_0(\mathbf{k})}{\partial k^\alpha\partial\mathbf{k}}\right)_{12}e^{i\omega t}&\partial ^\alpha E_1
    \end{array} \right)+\\\\
    e^2\mathrm{Re}\left[\frac{E^\alpha E^{*\beta}}{\hbar^2\omega^2}\right]\left(\begin{array}{cc}
         \left(\frac{\partial^3 H_0(\mathbf{k})}{\partial k^\alpha\partial k^\beta\partial k^\lambda}\right)_{22}& 0 \\
         0& \left(\frac{\partial^3 H_0(\mathbf{k})}{\partial k^\alpha\partial k^\beta\partial k^\lambda}\right)_{11}
    \end{array}\right),
\end{multline}

For future comparison let us write the density matrix derived from the Keldysh-Floquet approach ($-iG^<$) in the Schrödinger picture:
\begin{multline}\label{KFdensity matrixSI}
    \hat{\rho}_{\mathbf{K-F}}(t) = \left(\begin{array}{cc}
         f_2&0  \\
         0&f_1 
    \end{array}\right)+\\+\frac{f_1-f_2}{2(h^2+\frac{\Gamma^2}{4})}\left(\begin{array}{cc}
         h_x^2+h_y^2&h_+(h_z-i\frac\Gamma2)e^{-i\omega t}  \\\\h_-(h_z+i\frac\Gamma2)e^{i\omega t}
         &-h_x^2-h_y^2 
    \end{array}\right),
\end{multline}
where $h_{\pm} = h_x\pm ih_y$.

\subsection{The DC current in a clean limit}\label{secF}
In this section we will focus on the clean limit of the current Eqs.(\ref{kfcur1}-\ref{kfcur3}). If the electric field and $\Gamma$ are small, we can use the following approximation:
\begin{equation}
    \frac{1}{h^2+\frac{\Gamma^2}{4}}\approx \frac{2\pi\delta_\Lambda (\epsilon_{12}+\omega)}{\Lambda},
\end{equation}
where $\Lambda = \sqrt{|\mathbf{E}\cdot \mathbf{A}_{12}|^2+\frac{\Gamma^2}{4}}$ is the effective width of the Lorentzian and $\epsilon_{12}=\epsilon_1-\epsilon_2$.

By noticing that:
\begin{gather}
    h_yv^\alpha_x-v^\alpha_yh_x = \mathrm{Im}\left[(h_x+ih_y)(v^\alpha_x-iv^\alpha_y)\right],\\ h_xv^\alpha_x+v^\alpha_yh_y = \mathrm{Re}\left[(h_x+ih_y)(v^\alpha_x-iv^\alpha_y)\right],
\end{gather}
we can write this as:
\begin{multline}
    (h_x+ih_y)(v^\alpha_x-iv^\alpha_y)=\\=A^\beta A^{*\gamma}v^\gamma_{21}\left(\frac{\partial v^\beta}{\partial k^\alpha}\right)_{12}=\frac{E^\beta E^{*\gamma}}{\omega^2}v^\gamma_{21}\left(\frac{\partial v^\beta}{\partial k^\alpha}\right)_{12},
\end{multline}
where:
\begin{equation}
    \left(\frac{\partial v^\beta}{\partial k^\alpha}\right)_{12}=\partial^\alpha v^\beta_{12}+i[v^\beta,A^\alpha]_{12}.
\end{equation}

Now we can write the components of the current ($J_1,J_2,J_3$) as:
\begin{multline}
    J_1^\gamma = \pi \sum_{\mathbf{k}}\frac{\Gamma}{\sqrt{|\mathbf{E}\cdot \mathbf{A}_{12}|^2+\frac{\Gamma^2}{4}}}(f_1-f_2)\times\\\times \delta_\Lambda (\omega+\epsilon_{1}-\epsilon_2)  \mathrm{Im}\left[\frac{E^\alpha E^{*\beta}}{\omega^2}v^\beta_{21}\left(\frac{\partial v^\alpha}{\partial k^\gamma}\right)_{12}\right],
\end{multline}
\begin{multline}
    J_2^\gamma = \sum_{\mathbf{k}}\frac{\frac12(\omega+\epsilon_{1}-\epsilon_2)(f_1-f_2)}{\frac14(\omega+\epsilon_{1}-\epsilon_2)^2+\frac{(\epsilon_1-\epsilon_2)^2}{\omega^2}|\mathbf{E}\cdot \mathbf{A}_{12}|^2+\Gamma^2/4}\times\\\times\mathrm{Re}\left[\frac{E^\alpha E^{*\beta}}{\omega^2}v^\beta_{21}\left(\frac{\partial v^\alpha}{\partial k^\gamma}\right)_{12}\right],
\end{multline}
\begin{multline}
    J_3^\gamma = \pi \sum_{\mathbf{k}}(f_2-f_1)(v^\gamma_1-v^\gamma_2)\times\\\times\frac{|\mathbf{E}\cdot \mathbf{A}_{12}|^2\delta_\Lambda (\omega+\epsilon_{1}-\epsilon_2)}{\sqrt{|\mathbf{E}\cdot \mathbf{A}_{12}|^2+\frac{\Gamma^2}{4}}}.
\end{multline}

On resonance ($\omega\approx\epsilon_{12}$) we can use the following approximation:
\begin{equation}
    \frac{v^\beta_{21}}{\omega^2}\left(\frac{\partial v^\alpha}{\partial k^\gamma}\right)_{12}\approx A^\beta_{21}(\partial^\gamma A^\alpha_{12}-iA^\alpha_{12}(A^\gamma_{11}-A^\gamma_{22})).
\end{equation}

The expression above combined with the identity:
\begin{multline}
    \partial^\alpha A^\beta_{nm}-i[A^\alpha,\bar{A}^\beta]_{nm}=\\=\partial^\beta A^\alpha_{nm}-iA^\alpha_{nm}(A^\beta_{nn}-A^\beta_{mm}),
\end{multline}
recovers the perturbation results Eqs.(\ref{psr}-\ref{pir}).
The identification is the following: $J_1$ is the resonant shift current, $J_2$ is the non-resonant shift current, $J_3$ is the injection current, that in the clean limit ($\Gamma = 0$) are:
\begin{equation}
    J_1^\gamma = 0,
\end{equation}
\begin{multline}
    J_2^\gamma =e\sum_{\mathbf{k}}\frac{\frac12(\omega+\epsilon_{1}-\epsilon_2)(f_1-f_2)}{\frac14(\omega+\epsilon_{1}-\epsilon_2)^2+\frac{(\epsilon_1-\epsilon_2)^2}{\omega^2}|\mathbf{E}\cdot \mathbf{A}_{12}|^2}\times\\\times\mathrm{Re}\left[\frac{E^\alpha E^{*\beta}}{\omega^2}v^\beta_{21}\left(\frac{\partial v^\alpha}{\partial k^\gamma}\right)_{12}\right]
\end{multline}
\begin{equation}
    J_3^\gamma=e\sum_{\mathbf{k}}\frac{(f_2-f_1)(v^\gamma_1-v^\gamma_2)\frac{(\epsilon_1-\epsilon_2)^2}{\omega^2}|\mathbf{E}\cdot \mathbf{A}_{12}|^2}{(\omega+\epsilon_{1}-\epsilon_2)^2+4\frac{(\epsilon_1-\epsilon_2)^2}{\omega^2}|\mathbf{E}\cdot \mathbf{A}_{12}|^2}.
\end{equation}

\subsection{Periodic Gibbs Ensemble}\label{secG}
In the spirit of the rotating-wave approximation, and in order to consider the same level of approximation in which Keldysh-Floquet formalism is developed, we take the evolution is defined by the following truncated Hamiltonian:
\begin{equation}
    H^\mathbf{F}_{\mathrm{T}} = 
 \left(\begin{array}{cc}\epsilon_1+\omega&i(\mathbf{A}\cdot \mathbf{v}_{12})\\ 
 -i(\mathbf{A}^*\cdot \mathbf{v}_{21})&\epsilon_2\end{array}\right) = h_0+\mathbf{h}\cdot\mathbf{\sigma}.
\end{equation}

We can parametrize Hamiltonian vector in spherical coordinates, namely $\mathbf{h} = h (\cos\varphi\sin\theta,\sin\varphi\sin\theta,\cos\theta)$. 
Using the Hamiltonian above one can find the evolution operator in the Schrödinger picture:
\begin{gather}\label{evolution}
    U(t_{\mathrm{fin}},t_{\mathrm{in}}) = \psi_v(t_{\mathrm{fin}})\psi^\dagger_v(t_{\mathrm{in}})+\psi_c(t_{\mathrm{fin}})\psi^\dagger_c(t_{\mathrm{in}}),
\end{gather}
\begin{multline}
        \psi_v(t_{\mathrm{fin}})\psi^\dagger_v(t_{\mathrm{in}})=e^{iE_v(t_{\mathrm{in}}-t_{\mathrm{fin}})}\times\\\left(\begin{array}{cc}
         \cos^2\frac\theta2&-\sin\frac\theta2\cos\frac\theta2 e^{i\varphi-i\omega t_{\mathrm{in}}}  \\
         -\sin\frac\theta2\cos\frac\theta2 e^{-i\varphi+i\omega t_{\mathrm{fin}}}&\sin^2\frac\theta2 e^{-i\omega(t_{\mathrm{in}}-t_{\mathrm{fin}})} 
    \end{array}\right),
\end{multline}
\begin{multline}
        \psi_c(t_{\mathrm{fin}})\psi^\dagger_c(t_{\mathrm{in}})=e^{iE_c(t_{\mathrm{in}}-t_{\mathrm{fin}})}\times\\\left(\begin{array}{cc}
         \sin^2\frac\theta2&\sin\frac\theta2\cos\frac\theta2 e^{i\varphi-i\omega t_{\mathrm{in}}}  \\
         \sin\frac\theta2\cos\frac\theta2 e^{-i\varphi+i\omega t_{\mathrm{fin}}}&\cos^2\frac\theta2 e^{-i\omega(t_{\mathrm{in}}-t_{\mathrm{fin}})} 
    \end{array}\right),
\end{multline}
where the first argument of the evolution operator is the final time and the second - the initial one:
\begin{equation}
    \psi(t_{\mathrm{fin}}) = U(t_\mathrm{fin},t_\mathrm{in})\psi(t_\mathrm{in}).
\end{equation}

Now we follow the PGE procedure \cite{Moess1,Moess2,Moess3,Moess4}. The system of interest has a conserved quantity that is the total number of particles. We construct the PGE density matrix in a way that it conserves the given quantity during the evolution. Initially occupations are given by:
\begin{gather}  
\psi_v(0)\psi^\dagger_v(0) = \left(\begin{array}{cc}
         \cos^2\frac\theta2& -\frac12\sin\theta e^{i\varphi} \\
         -\frac12\sin\theta e^{-i\varphi}&\sin^2\frac\theta2 
    \end{array}\right),\\
    \psi_c(0)\psi^\dagger_c(0) = \left(\begin{array}{cc}
         \sin^2\frac\theta2& \frac12\sin\theta e^{i\varphi} \\
         \frac12\sin\theta e^{-i\varphi}&\cos^2\frac\theta2 
    \end{array}\right).
\end{gather}

The time evolution of occupation numbers above is:
\begin{gather}
    \mathcal{I}_{v/c}(t) = U(0,t)\psi_{v/c}(0)\psi^\dagger_{v/c}(0) U^\dagger(0,t),\\
    \mathcal{I}_c(t)+\mathcal{I}_v(t)=1.\label{secondeqau}
\end{gather}

Now we can can construct the PGE density matrix that satisfies the initial condition:
\begin{gather}
    \hat{\rho}_{\mathrm{PGE}}(t)= \mathcal{Z}^{-1}\exp\left(-\lambda_c\mathcal{I}_c(t)-\lambda_v\mathcal{I}_v(t)\right),\label{pgedensity matrix}\\
    \mathrm{Tr}\left[\rho_\mathrm{S}(0)\mathcal{I}_{v/c}(0)\right] = \mathrm{Tr}\left[\hat{\rho}_{\mathrm{PGE}}(0)\mathcal{I}_{v/c}(0)\right],\label{initicond}
\end{gather}

The initial state of the system is chosen to be the thermal state (see the discussion in the main text) with the temperature and chemical potential of the bath:
\begin{gather}
     \rho_\mathrm{S}(0) = \left(\begin{array}{cc}
         f_2&0  \\
         0&f_1
    \end{array}\right).
\end{gather}

Solving Eq.(\ref{initicond}) one can obtain:
\begin{equation}
     \lambda_c-\lambda_v = \ln \left[\frac{1}{f_2-(f_2-f_1)\cos^2\frac\theta2}-1\right].\label{lambdaeq}
\end{equation}

Using Eq.(\ref{secondeqau}) and Eq.(\ref{lambdaeq}) one can rewrite the PGE density matrix from Eq.(\ref{pgedensity matrix}) as:
\begin{multline}
    \hat{\rho}_{\mathbf{PGE}}(t)= \left(\begin{array}{cc}
         \frac12&0  \\
         0& \frac12
    \end{array}\right)+\\\left(\begin{array}{ccc}
         \frac{(f_2-f_1)}{2}\cos^2\theta&\frac{(f_1-f_2)}{2}\sin\theta\cos\theta e^{i\varphi-i\omega t}  \\\\\frac{(f_1-f_2)}{2}\sin\theta\cos\theta e^{-i\varphi+i\omega t}
         & -\frac{(f_2-f_1)}{2}\cos^2\theta
    \end{array}\right).
\end{multline}

Using the following relations:
\begin{gather}
    \sin\theta e^{i\varphi} = \frac{h_x+ih_y}{h}=\frac{h_+}{h},\qquad
    \sin\theta e^{-i\varphi}=\frac{h_-}{h},\\  \cos\theta = \frac{h_z}{h},\qquad h^2 = h_x^2+h_y^2+h_z^2,
\end{gather}
one can show that the PGE density matrix is:
\begin{multline}\label{PGEsupl}
    \hat{\rho}_{\mathbf{PGE}}(t)= \left(\begin{array}{cc}
         f_2&0  \\
         0& f_1
    \end{array}\right)+\\+\frac{f_1-f_2}{2h^2}\left(\begin{array}{ccc}
         h_x^2+h_y^2&h_z h_+ e^{-i\omega t}  \\\\h_zh_-e^{i\omega t}
         & -h_x^2-h_y^2
    \end{array}\right).
\end{multline}
Which is the same matrix introduced in Eq.(12) of the main text.

\subsection{Rabi oscillations}\label{secH}
We consider the 2 band system with the evolution operator from Eq.(\ref{evolution}). 
We assume the system initially to be thermal with the temperature and chemical potential of the bath, namely:
\begin{gather}
     \rho_\mathrm{S}(0) = \left(\begin{array}{cc}
         f_2&0  \\
         0&f_1
    \end{array}\right).
\end{gather}

One can show that the evolution of the conduction  and valence bands are given by:
\begin{gather}
    \ket{v(t)}= e^{-ih_0t}\left(\begin{array}{c}
           -i\sin\theta\sin(ht)e^{i\varphi}\\
          (\cos(ht)-i\cos\theta\sin(ht))e^{i\omega t}
    \end{array}\right),\\\nonumber\\
    \ket{c(t)}= e^{-ih_0t}\left(\begin{array}{c}
           \cos(ht)+i\cos\theta\sin(ht)\\
          -i\sin\theta\sin(ht)e^{i\omega t-i\varphi}
    \end{array}\right),
\end{gather}

Now we can construct the Rabi density matrix as follows:
\begin{equation}
    \rho_{\mathbf{Rabi}}(t) = f_1\ket{v(t)}\bra{v(t)}+f_2\ket{c(t)}\bra{c(t)}.
\end{equation}

To phenomenologically capture the  synchronization of the system with drive, we perform a time average of the terms that have a frequencies different from the drive frequency $\omega$, in the above the density matrix, as follows:
\begin{gather}
    \overline{\sin(ht)^2} = \frac12,\qquad \overline{\cos(ht)^2} = \frac12,\qquad
    \overline{\sin(ht)} = 0,\\  \overline{\cos(ht)} = 0,\qquad
     \overline{\sin(ht)\cos(ht)} = 0.
\end{gather}
Which leads to the following synchronized Rabi density matrix:
\begin{multline}
    \hat{\rho}_{\mathbf{Rabi}}(t)= \left(\begin{array}{cc}
         \frac12&0  \\
         0& \frac12
    \end{array}\right)+\\\left(\begin{array}{ccc}
         \frac{(f_2-f_1)}{2}\cos^2\theta&\frac{(f_1-f_2)}{2}\sin\theta\cos\theta e^{i\varphi-i\omega t}  \\\\\frac{(f_1-f_2)}{2}\sin\theta\cos\theta e^{-i\varphi+i\omega t}
         & -\frac{(f_2-f_1)}{2}\cos^2\theta
    \end{array}\right),
\end{multline}
which also can be simplified to:
\begin{multline}
    \hat{\rho}_{\mathbf{Rabi}}(t)= \left(\begin{array}{cc}
         f_2&0  \\
         0& f_1
    \end{array}\right)+\\+\frac{f_1-f_2}{2h^2}\left(\begin{array}{ccc}
         h_x^2+h_y^2&h_z h_+ e^{-i\omega t}  \\\\h_zh_-e^{i\omega t}
         & -h_x^2-h_y^2
    \end{array}\right).
\end{multline}
Which is the same matrix as that obtained from the PGE in Eq.(\ref{PGEsupl}) and introduced in Eq.(12) of the main text.
\subsection{Injection current for 3D Weyl and 2D Dirac Fermions}\label{analitic}

Here we derive the approximate analytic expression of the injection current. The resonant injection current is given by:
\begin{multline}\label{hinj}
    {J}^\alpha_{3}=\pi  \frac{e}{\hbar}\int\frac{d\mathbf{k}}{(2\pi)^3}(f_2-f_1)(v^\alpha_1-v^\alpha_2)\times\\\times\frac{\left|e\mathbf{E}\cdot \mathbf{A}_{12}\right|^2}{\sqrt{\left|e\mathbf{E}\cdot \mathbf{A}_{12}\right|^2+\frac{\Gamma^2}{4}}}\delta(\epsilon_1-\epsilon_2 +\hbar\omega).
\end{multline}

In the $T\rightarrow0$ limit, the valence/conduction band occupation difference is $f_1-f_2 = \Theta(\epsilon_c-\epsilon_F)$, where $\Theta$ is the Heaviside theta function and $\epsilon_F$ is the Fermi energy.

\subsubsection{3D Weyl fermions}

The Hamiltonian of 3D Weyl fermions is given by:
\begin{equation}\label{3DWs}
    \hat{H}_0 = \sum_{\alpha=x,y,z}{k_\alpha}\cdot{\hat{\sigma}_\alpha},
\end{equation}
here we have re-scaled the momentum $v_0 \mathbf{k}\rightarrow \mathbf{k}, $ to simplify the final expression. The components of the momentum vector are $\mathbf{k} = k \mathbf{n}= k(\cos\phi\sin\theta,\sin\phi\sin\theta,\cos\theta)$. Consequently, the off-diagonal Berry connections and band energy difference of the Hamiltonian Eq.(\ref{3DWs}) are given by:
\begin{gather}
    A^{x}_{12}=\frac{\sin\phi}{2k}-i\frac{\cos\phi\cos\theta}{2k},\quad A^{y}_{12}=-\frac{\cos\phi}{2k}-i\frac{\sin\phi\cos\theta}{2k}\\A^{z}_{12}=i\frac{\sin\theta}{2k},\quad
    \epsilon_{2}-\epsilon_1 = 2k, \quad v^\gamma_2-v^\gamma_1 = 2n^\gamma. 
\end{gather}

Note that $\mathbf{A}^*_{12}=\mathbf{A}_{21}$. The direction of the injection current behaves as  $\mathbf{J}_3\sim \left[\mathbf{E}\times\mathbf{E}^*\right]$. We consider the frequency of the drive to be $\omega > 2\epsilon_F$, which sets $f_1-f_2 = 1$. The injection current can be then approximated as: 
\begin{equation}\label{eq148}
    J_3^\gamma =\frac{\pi\omega}{\sqrt{|\mathbf{E}|^2+\frac{\Gamma^2\omega^2}{4}}}\int \frac{|\mathbf{E}\cdot \mathbf{A}_{12}|^2\delta_\Lambda (\omega-2k)}{\sqrt{1 +\frac{\omega^2|\mathbf{E}\cdot \mathbf{A}_{12}|^2-|\mathbf{E}|^2}{{|\mathbf{E}|^2+\frac{\Gamma^2\omega^2}{4}}}}}\frac{n^\gamma d^3\mathbf{k}}{(2\pi)^3}.
\end{equation}
The term containing the square root inside the integral is of the form $(1+X)^{-1/2}$, with $|X|<1$. We therefore expand this term as $(1+X)^{-1/2}\approx 1-X/2$, and obtain:
\begin{equation}
\mathbf{J}_3\approx \frac{\pi^2 i \omega}{(2\pi)^3} \frac{ \left[\mathbf{E}^*\times\mathbf{E}\right]}{60\sqrt{|\mathbf{E}|^2+\frac{\Gamma^2\omega^2}{4}}} \frac{12|\mathbf{E}|^2+5\Gamma^2\omega^2 }{|\mathbf{E}|^2+\frac{\Gamma^2\omega^2}{4}},
\end{equation}
which after units restoring is:
\begin{multline}\label{appj3rl}
    \mathbf{J}_3 \approx \frac{i\pi^2 e^2 \omega}{v_0(2\pi)^3} \frac{ \left[\mathbf{E}^*\times\mathbf{E}\right]}{60\sqrt{|\mathbf{E}|^2+\frac{\Gamma^2\hbar^2\omega^2}{4v_0^2e^2}}} \frac{12|\mathbf{E}|^2+5\frac{\Gamma^2\hbar^2\omega^2}{v_0^2e^2} }{|\mathbf{E}|^2+\frac{\Gamma^2\hbar^2\omega^2}{4v_0^2e^2}}.
\end{multline}
The above expansion can only be justified parametrically either when $\Gamma \hbar\omega \gg ev_0|\mathbf{E}|$ or when the light is almost linearly polarized at arbitrary $\Gamma$, however, as we will show by explicit numerical evaluation of the integral, the approximation still works to about $17\%$ even for perfectly circularly polarized light. 

If the electric field is perfectly circularly polarised, the injection current of ideal Weyl model can be evaluated from the full integral in Eq.(\ref{hinj}) and yields the following expression:
\begin{multline}\label{exactj3rl}
    \mathbf{J}_3 =\frac{ie^3[\mathbf{E}^*\times\mathbf{E}]}{12\pi\Gamma\hbar}\times\\\times\frac{8-8\sqrt{\mathcal{E}^2+1}+\mathcal{E}^2\sqrt{1+\mathcal{E}^2}+3\mathcal{E}\mathrm{ArcSinh}(\mathcal{E})}{\mathcal{E}^2},
\end{multline}
where $\mathcal{E}=2\sqrt{2}\frac{e |\mathbf{E}|v_0}{\hbar \omega\Gamma}$. This expression is in fact exactly identical to that derived in Ref.\cite{Leppenen2019} in the limit of $\tau_\epsilon/\tau_p \rightarrow 0$, where $\tau_{\epsilon,p}$ are the energy and momentum relaxation times introduced in Ref.\cite{Leppenen2019}. The expression above in the limit of a large electric field $\mathcal{E}\rightarrow\infty$ can be approximated as:
\begin{equation}
    \mathbf{J}^{\rm exact}_3 \approx\frac{1}{24\sqrt{2}\pi}[\mathbf{E}^*\times\mathbf{E}]\frac{ie^2\omega}{|\mathbf{E}|v_0}.
\end{equation}
On the other hand the approximation from Eq.(\ref{appj3rl}) gives:
\begin{figure}[t]\label{sibfig1}
\centering
\includegraphics[width=8cm]{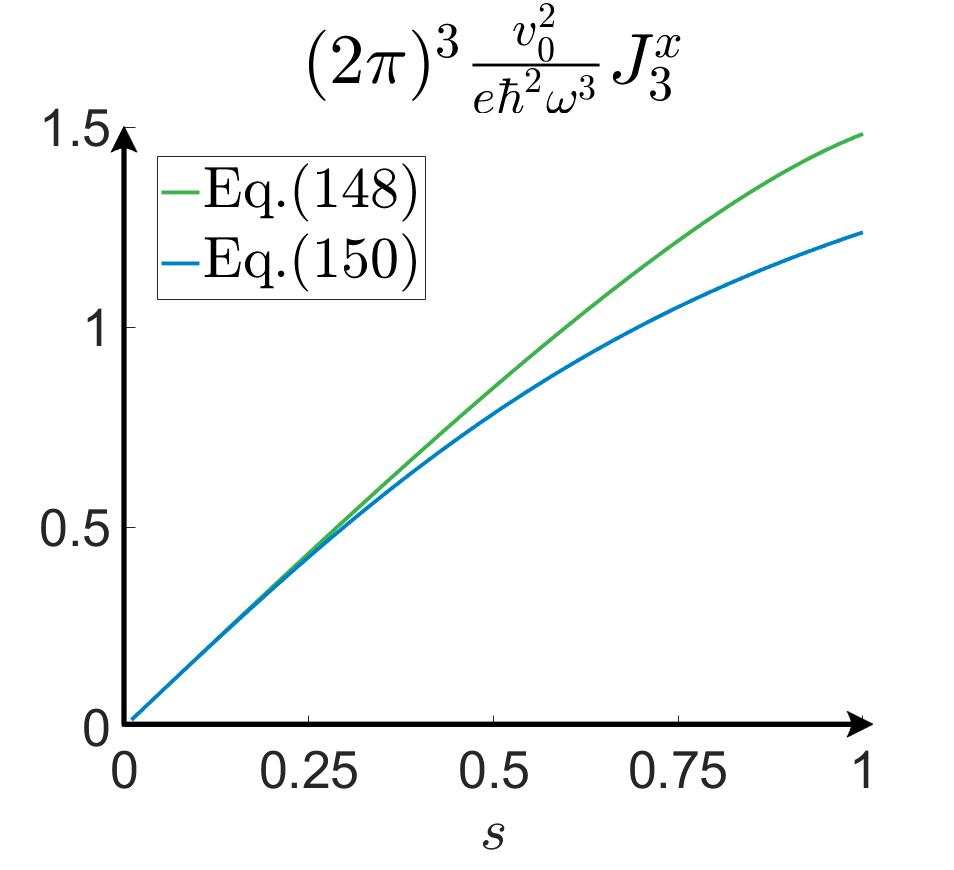}
\caption{Comparison of exact evaluation of the integral from Eq.(\ref{eq148}) and the approximate expression from Eq.(\ref{appj3rl}) for the injection current for 3D Weyl fermion as a function of a polarisation $s$, such that $s\rightarrow0$ for linearly polarizes light and $s\rightarrow1$ for perfectly circularly polarized light (see Eq.\ref{eq154}) in the Rabi regime $ev_0|\mathbf{E}| \gg \Gamma\hbar\omega $.}
\end{figure}
\begin{equation}
    \mathbf{J}^{\rm approx}_3\approx \frac{1}{40\pi}[\mathbf{E}^*\times\mathbf{E}]\frac{ie^2\omega}{|\mathbf{E}|v_0}.
\end{equation}
Despite the fact that we see a small discrepancy between approximation and exact calculation ($\mathbf{J}^{\text{exact}}_3/\mathbf{J}^{\rm approx}_3\approx 1.17$), formula Eq.(\ref{appj3rl}) displays good agreement with the exact evaluation of the integral in Eq.(\ref{eq148}) for light that is not perfectly circularly polarized. To illustrate this we consider light with elliptical polarization that interpolates from perfectly linearly polarized to perfectly circularly polarized as follows:
\begin{equation}\label{eq154}
    \mathbf{E} = (0, i s \frac{E_0}{\sqrt{2}},\frac{E_0}{\sqrt{2}}),\qquad s\in[0,1].
\end{equation}
Parameters are chosen outside of the regime in which the approximation of Eq.(\ref{appj3rl}) is expected to be justified, namely $ev_0|\mathbf{E}| \gg \Gamma\hbar\omega$ ($\hbar\omega = 2.5\epsilon_F, \Gamma = 0.01\hbar\omega, ev_0E_0 = 10\Gamma\hbar\omega$). The result of comparison can be seen on FIG.\ref{sibfig1}. We see that Eq.(\ref{appj3rl}) matches the exact integral from Eq.(\ref{eq148}) for light that is almost linearly polarized ($s\approx 0$) and deviates from it the most in the case of perfect circularly polarized light ($s=1$), but only by about $17\%$. Therefore, we conclude that Eq.(\ref{appj3rl}) produces a good approximation of the photocurrent current over different regimes.  
\subsubsection{2D Dirac fermions}
After the analogous momentum rescaling the Hamiltonian of 2D tilted Dirac fermions is given by:
\begin{equation}
    \hat{H} = \alpha k_x \hat{\mathds{1}} + k_x\hat{\sigma}_x+k_y\hat{\sigma}_y +m\hat{\sigma}_z,
\end{equation}
where $\alpha = u_x/v_x$. The momentum vector is two dimentional $\mathbf{k} = k\mathbf{n} =k(\cos\phi,\sin\phi)$. The off-diagonal berry connections, band energies and velocity differences are:
\begin{gather}
    A^x_{12} = \frac{\sqrt{k^2+m^2}\sin\phi+im\cos\phi}{2(k^2+m^2)},\\
    A^y_{12} =- \frac{\sqrt{k^2+m^2}\cos\phi-im\sin\phi}{2(k^2+m^2)},\\
    \epsilon_{2}-\epsilon_1 = 2 \sqrt{k^2+m^2},\quad
    \mathbf{v}_2 -\mathbf{v}_1 = \frac{2k}{\sqrt{k^2+m^2}} \mathbf{n}.
\end{gather}
Assuming $\omega = 2\epsilon_F$, we see that $f_1-f_2 = \Theta(\alpha k\cos \phi)\rightarrow \phi \in [-\pi/2,\pi/2]$. For simplicity we assume that the electric field is along the tilt so that the corresponding component of the injection current can be approximated as:
\begin{multline}
    J_3^x =\frac{ k_0}{2\pi}\frac{1}{\sqrt{\frac{|\mathbf{E}|^2}{\epsilon_F^2}+\Gamma^2}}\times\\\times\int_{-\pi/2}^{\pi/2}\int_0^{\infty}  n^x\frac{|(\mathbf{E}\cdot\mathbf{A}_{12})|^2}{\sqrt{1+\frac{{4|(\mathbf{E}\cdot\mathbf{A}_{12})|^2 - \frac{|\mathbf{E}|^2}{\epsilon_F^2}}}{\frac{|\mathbf{E}^2|}{\epsilon_F^2}+\Gamma^2}}}\delta(k-k_0)dkd\phi\approx\\\approx \frac{\sqrt{\epsilon_F^2-m^2}}{60\pi}\frac{\frac{|\mathbf{E}|^2}{\epsilon_F^2}}{\sqrt{\frac{|\mathbf{E}|^2}{\epsilon_F^2}+\Gamma^2}}\times\\\times\frac{5\Gamma(1+2\frac{m^2}{\epsilon_F^2})+\frac{|\mathbf{E}|^2}{\epsilon_F^2}(6+13\frac{m^2}{\epsilon_F^2}-4\frac{m^4}{\epsilon_F^4})}{\frac{|\mathbf{E}|^2}{\epsilon_F^2}+\Gamma^2},
\end{multline}
which after units restoring is:
\begin{multline}
    J_3^x \approx \frac{e^2v_x}{\hbar v_y}\frac{\sqrt{\epsilon_F^2-m^2}}{60\pi}\frac{\frac{|\mathbf{E}|^2}{\epsilon_F^2}}{\sqrt{\frac{|\mathbf{E}|^2}{\epsilon_F^2}+\frac{\Gamma^2}{e^2v_x^2}}}\times\\\times\frac{5\frac{\Gamma^2}{e^2v^2_x}(1+2\frac{m^2}{\epsilon_F^2})+\frac{|\mathbf{E}|^2}{\epsilon_F^2}(6+13\frac{m^2}{\epsilon_F^2}-4\frac{m^4}{\epsilon_F^4})}{\frac{|\mathbf{E}|^2}{\epsilon_F^2}+\frac{\Gamma^2}{e^2v_x^2}}.
\end{multline}
\subsection{The shift current for linear in momentum models}\label{supshift}
This section focuses on contributions of shift current in models that are linear in momentum $\mathbf{k}$. We demonstrate why the contribution of shift currents is neglible in such models and the rectification response is dominated by the the injection current in the limit where the two band approximation is valid. 

\subsubsection{Two-band Keldysh-Floquet formalism}
In the Keldysh-Floquet formalism employed in the main text, the velocity operator from Eq.(\ref{velocityfloq}) for the linear in momentum model simplifies to the following expression:
\begin{equation}
    v^{\alpha}_{\mathbf{F}} = \left(\begin{array}{cc}
         \partial^\alpha \epsilon_1& 0 \\
         0&\partial ^\alpha \epsilon_2
    \end{array} \right),
\end{equation}
which is a time-independent operator in Schrödinger's picture. Therefore, we see that there are no associated off-diagonal components of the velocity operator, which are the ones that would give rise to shift currents. After averaging of the density matrix from Eq.(4) of the main text with the velocity operator written above, the resultant expression of the current picks only the contribution from the injection current:
\begin{multline}
    J^\alpha = -i\mathbf{Tr}\left[\hat{G}^<_{\mathbf{F}}\hat{v}^\alpha_{\mathbf{F}}\right] =\\= -\int\frac{d\mathbf{k}}{(2\pi)^3}(f_1-f_2)v^\alpha_z\frac{h_x^2+h_y^2}{h^2+\frac{\Gamma^2}{4}}= J_3^\alpha,
\end{multline}
which means that in the two-band approximation of the linear in momentum model $\mathbf{k}$ the shift current is absent.  

\subsubsection{3D Weyl model's shift current from the perturbation theory}
Now, let's consider the prediction of the perturbation theory for the 3D Weyl fermion model, which Hamiltonian is:
\begin{equation}\label{3dWsup}
    H = \boldsymbol{\sigma}\cdot \boldsymbol{k}.
\end{equation}
According to the perturbation theory, the following expression gives the shift current part of the rectification conductivity:
\begin{multline}\label{shift3dwsup}
    \sigma^{\gamma\beta\alpha}_{\mathrm{S}}(-\omega,\omega) = \frac12\int\frac{d\mathbf{k}}{(2\pi)^3}\times\\\times\sum_{nm}\Bigg\{ \underbrace{\hat{A}^{\gamma}_{mn}\frac{\partial}{\partial k^\alpha}\frac{(f_{n}-f_m)\hat{A}^{\beta}_{nm}}{\omega-\epsilon_{n}+\epsilon_{m}+i\Gamma}}_{\text{contribution 1}} 
    +\\+ \underbrace{i\frac{(f_{n}-f_m)\hat{A}^{\beta}_{nm}}{\omega  -\epsilon_{n}+\epsilon_{m}+i\Gamma}\sum_c \bigg[\hat{A}^\alpha_{mc}\hat{\bar{A}}^{\gamma}_{cn}-\hat{\bar{A}}^{\gamma}_{mc}\hat{A}^\alpha_{cn}\bigg]}_{\text{contribution 2}}\Bigg\}+\\+\left(\begin{array}{c}\alpha\leftrightarrow \beta\\\omega\leftrightarrow-\omega\end{array}\right).
\end{multline}
Both contributions can be analytically computed within this model, and are given by:
\begin{multline}\label{supcontr1}
    \sigma^{\gamma\beta\alpha}_{\text{contribution 1}}(\omega)=\frac{2}{(2\pi)^3}\bigg(\Phi^{\gamma\beta\alpha}(I_{11}(\omega)+c.c.)+\\+\frac{i\pi\Gamma\varepsilon^{\gamma\beta\alpha}}{6}(I_{12}(\omega)-c.c.)\bigg),
\end{multline}
\begin{equation}\label{supcontr2}
    \sigma^{\gamma\beta\alpha}_{\text{contribution 2}}(\omega)=\frac{-2}{(2\pi)^3}\Phi^{\gamma\beta\alpha}(I_{11}(\omega)+c.c.),
\end{equation}
where
\begin{equation}
    I_{ab}(\omega)=i\int_{\epsilon_F}^{\infty}\frac{dE}{(\omega-2E+i\Gamma)^{a}(-2E+i\Gamma)^{b}}
\end{equation}
and 
\begin{gather}
    \Phi^{x\beta\alpha}=\left(\begin{array}{ccc} 0&0&0\\0&0&-\frac\pi3 \\0&-\frac\pi3&0 \end{array}\right),\\\Phi^{y\beta\alpha}=\left(\begin{array}{ccc} 0&0&\frac\pi3\\0&0&0 \\\frac\pi3&0&0 \end{array}\right),\qquad\Phi^{z\beta\alpha}=\left(\begin{array}{ccc} 0&0&0\\0&0&0 \\0&0&0 \end{array}\right).
\end{gather}
After adding these two contributions we obtain:
\begin{equation}
    \sigma^{\gamma\beta\alpha}_{\mathrm{S}}(-\omega,\omega) =\frac{i\pi\Gamma\varepsilon^{\gamma\beta\alpha}}{3(2\pi)^3}(I_{12}(\omega)-c.c.).
\end{equation}
As one can see, the shift current contribution of 3D ideal Weyl fermion is of order $\Gamma$ and can be neglected in a clean limit. 

Alternatively, one could use the time-reversal and rotational symmetry argument to show that the perturbative shift current vanishes for the Hamiltonian of Eq.(\ref{3dWsup}) in the clean limit ($\Gamma\rightarrow0$). First, the shift conductivity from Eq.(\ref{shift3dwsup}) of Hamiltonian Eq.(\ref{3dWsup}) has to be a three index tensor, that is symmetric under $SO(3)$ transformations. $\pi$ rotational symmetry around $x,y,z$ axes limits only those components of the tensor Eq.(\ref{shift3dwsup}) to be finite, which all three indices are distinct. Additionally, the tensor must be symmetric under cyclic permutations of $(x,y,z)$ labels, as these can be implemented as a subgroup of $SO(3)$. These properties force the tensor to be proportional to the Levi-Civita tensor, and we conclude that the shift conductivity from Eq.(\ref{shift3dwsup}) of Hamiltonian Eq.(\ref{3dWsup}) has to be proportional to it. On the other hand, it can be shown that in the clean limit the resonant part of the shift current is even under time reversal symmetry (see table 1 of Ref.\cite{ahn2020lowfrequency}). Nevertheless the part of bilinear of electric fields that contracts with the Levi-Civita tensor is its circularly polarized component, which is a time-reversal-odd, and thus such components must be absent from the perturbative expressions of the shift current in the clean limit~\cite{ahn2020lowfrequency}. In summary,  the combination of three ingredients: time-reversal symmetry, $SO(3)$ symmetry and the clean limit ($\Gamma\rightarrow0$), force the shift current of the ideal Weyl models to vanish.

\subsubsection{2D Dirac model's shift current from the perturbation theory}
Next, we focus on another linear momentum model - the 2D Dirac model. The Hamiltonian of this model is:
\begin{equation}
    \hat{H} = \frac{u_x}{v_x} k_x \hat{\mathds{1}} + k_x\hat{\sigma}_x+k_y\hat{\sigma}_y +m\hat{\sigma}_z.
\end{equation}
According to the perturbation theory, the resonant part of the shift conductivity is:
\begin{multline}
    \sigma^{\gamma\beta\alpha}_{\mathrm{S,R}}(-\omega,\omega) = \frac\pi2\frac{e^3}{\hbar^2}\int\frac{d\mathbf{k}}{(2\pi)^2}\delta(\omega-\epsilon_{n}+\epsilon_{m})\times\\\times\sum_{nm}\Bigg\{ (f_{n}-f_m)\hat{A}^{\beta}_{nm} 
    i\frac{\partial}{\partial k^\alpha}\hat{A}^{\gamma}_{mn}+\\+ (f_{n}-f_m)\hat{A}^{\beta}_{nm}\sum_c \bigg[\hat{A}^\alpha_{mc}\hat{\bar{A}}^{\gamma}_{cn}-\hat{\bar{A}}^{\gamma}_{mc}\hat{A}^\alpha_{cn}\bigg]\Bigg\}+\\+\left(\begin{array}{c}\alpha\leftrightarrow \beta\\\omega\leftrightarrow-\omega\end{array}\right).
\end{multline}
It turns out that in this model the shift current is finite. The value of the tilt, $u_x/v_x$, determines the frequency window of non-zero resonant shift current, and the maximum value of the conductivity for small tilts occurs near the middle of this window at a frequency $\omega \approx 2 \epsilon_F$ (see Fig.2(c) of the main text). The peak values of the conductivity can be analytically computed and are given by (here $\hbar=e=v_x=v_y=1$): 
\begin{equation}
    \sigma^{x\beta\alpha}_{\mathrm{S,R}}(-\omega,\omega)  =\frac{1}{8\pi}\left(\begin{array}{cc}
         0&\frac{i m\sqrt{\epsilon_F^2-m^2}}{\epsilon_F^4}  \\
         \frac{i m\sqrt{\epsilon_F^2-m^2}}{\epsilon_F^4}&0 
    \end{array}\right),
\end{equation}
\begin{equation}
    \sigma^{y\beta\alpha}_{\mathrm{S,R}}(-\omega,\omega) =\frac{1}{8\pi}\left(\begin{array}{cc}
         -2im\frac{\sqrt{\epsilon_F^2-m^2}}{\epsilon_F^4}&-\frac{ \sqrt{\epsilon_F^2-m^2}}{\epsilon_F^3}  \\
         \frac{ \sqrt{\epsilon_F^2-m^2}}{\epsilon_F^3}&0 
    \end{array}\right),
\end{equation}
whereas the injection part is:
\begin{multline}
    \sigma^{x\beta\alpha}_{\mathrm{I,R}}(-\omega,\omega)  =\\=\frac{1}{8\pi\Gamma}\left(\begin{array}{cc}
         \frac{\sqrt{\epsilon_F^2-m^2}(2m^2+\epsilon_F^2)}{3\epsilon_F^4}&\frac{i m\sqrt{\epsilon_F^2-m^2}}{\epsilon_F^3}  \\
         -\frac{i m\sqrt{\epsilon_F^2-m^2}}{\epsilon_F^3}&\frac{\sqrt{\epsilon_F^2-m^2}(2m^2+\epsilon_F^2)}{3\epsilon_F^4}
    \end{array}\right),
\end{multline}
\begin{equation}
    \sigma^{y\beta\alpha}_{\mathrm{I,R}}(-\omega,\omega) =\frac{1}{8\pi\Gamma}\left(\begin{array}{cc}
         0&-\frac{m \left(\sqrt{\epsilon_F^2-m^2}\right)^{3/2}}{3\epsilon_F^4}  \\
         -\frac{m \left(\sqrt{\epsilon_F^2-m^2}\right)^{3/2}}{3\epsilon_F^4}&0 
    \end{array}\right),
\end{equation}
We see that the resonant shift currents do not vanish exactly for linear in $\mathbf{k}$ models of 2D Dirac fermions from perturbation theory. However, the injection conductivity is typically parametrically larger than the shift conductivity in the limit in which one is justified to focus only on the two bands ($\Gamma^2\ll\epsilon_F^2-m^2\ll\epsilon_F^2$), which is the same regime in which our rotating-wave two-band truncation of the Floquet-Keldysh formalism is justified. In this limit the dominant contribution of the shift conductivity typically scales as $\sigma_{\mathrm{S,R}}\sim\sqrt{\epsilon_F^2-m^2}/\epsilon_F^3$, while the injection conductivity typically scales as $\sigma_{\mathrm{I,R}}\sim\sqrt{\epsilon_F^2-m^2}/(\epsilon_F^2\Gamma)$, and the ratio of them is $\sigma_{\mathrm{I,R}}/\sigma_{\mathrm{S,R}}\sim\epsilon_F/\Gamma$. Consequently, the shift current contribution can be neglected. 

\bibliography{mysuperbib}

\begin{thebibliography}{66}
\expandafter\ifx\csname natexlab\endcsname\relax\def\natexlab#1{#1}\fi
\expandafter\ifx\csname bibnamefont\endcsname\relax
  \def\bibnamefont#1{#1}\fi
\expandafter\ifx\csname bibfnamefont\endcsname\relax
  \def\bibfnamefont#1{#1}\fi
\expandafter\ifx\csname citenamefont\endcsname\relax
  \def\citenamefont#1{#1}\fi
\expandafter\ifx\csname url\endcsname\relax
  \def\url#1{\texttt{#1}}\fi
\expandafter\ifx\csname urlprefix\endcsname\relax\def\urlprefix{URL }\fi
\providecommand{\bibinfo}[2]{#2}
\providecommand{\eprint}[2][]{\url{#2}}

\bibitem[{\citenamefont{Belinicher and Sturman}(1980)}]{Belinicher1980}
\bibinfo{author}{\bibfnamefont{V.~I.} \bibnamefont{Belinicher}}
  \bibnamefont{and} \bibinfo{author}{\bibfnamefont{B.~I.}
  \bibnamefont{Sturman}}, \bibinfo{journal}{Phys. Usp.}
  \textbf{\bibinfo{volume}{23}}, \bibinfo{pages}{199} (\bibinfo{year}{1980}).

\bibitem[{\citenamefont{von Baltz}(1981)}]{BaltzN}
\bibinfo{author}{\bibfnamefont{R.}~\bibnamefont{von Baltz}},
  \bibinfo{journal}{Ferroelectrics} \textbf{\bibinfo{volume}{35}},
  \bibinfo{pages}{131–136} (\bibinfo{year}{1981}).

\bibitem[{\citenamefont{Sturman and Fridkin}(1992)}]{Sturman1992}
\bibinfo{author}{\bibfnamefont{B.~I.} \bibnamefont{Sturman}} \bibnamefont{and}
  \bibinfo{author}{\bibfnamefont{V.~M.} \bibnamefont{Fridkin}},
  \emph{\bibinfo{title}{The photovoltaic and photorefractive effects in
  noncentrosymmetric materials}}, no. \bibinfo{number}{v. 8} in
  \bibinfo{series}{Ferroelectricity and related phenomena}
  (\bibinfo{publisher}{Gordon and Breach Science Publishers},
  \bibinfo{address}{Philadelphia}, \bibinfo{year}{1992}).

\bibitem[{\citenamefont{Kraut and von Baltz}(1979)}]{PhysRevB.19.1548}
\bibinfo{author}{\bibfnamefont{W.}~\bibnamefont{Kraut}} \bibnamefont{and}
  \bibinfo{author}{\bibfnamefont{R.}~\bibnamefont{von Baltz}},
  \bibinfo{journal}{Phys. Rev. B} \textbf{\bibinfo{volume}{19}},
  \bibinfo{pages}{1548} (\bibinfo{year}{1979}).

\bibitem[{\citenamefont{von Baltz and Kraut}(1981)}]{PhysRevB.23.5590}
\bibinfo{author}{\bibfnamefont{R.}~\bibnamefont{von Baltz}} \bibnamefont{and}
  \bibinfo{author}{\bibfnamefont{W.}~\bibnamefont{Kraut}},
  \bibinfo{journal}{Phys. Rev. B} \textbf{\bibinfo{volume}{23}},
  \bibinfo{pages}{5590} (\bibinfo{year}{1981}).

\bibitem[{\citenamefont{Belinicher et~al.}(1982)\citenamefont{Belinicher,
  Ivchenko, and Sturman}}]{Belinicher}
\bibinfo{author}{\bibfnamefont{V.}~\bibnamefont{Belinicher}},
  \bibinfo{author}{\bibfnamefont{E.}~\bibnamefont{Ivchenko}}, \bibnamefont{and}
  \bibinfo{author}{\bibfnamefont{B.}~\bibnamefont{Sturman}},
  \bibinfo{journal}{Zh. Eksp. Teor. Fiz.} \textbf{\bibinfo{volume}{83}},
  \bibinfo{pages}{649} (\bibinfo{year}{1982}).

\bibitem[{\citenamefont{Aversa and Sipe}(1995)}]{PhysRevB.52.14636}
\bibinfo{author}{\bibfnamefont{C.}~\bibnamefont{Aversa}} \bibnamefont{and}
  \bibinfo{author}{\bibfnamefont{J.~E.} \bibnamefont{Sipe}},
  \bibinfo{journal}{Phys. Rev. B} \textbf{\bibinfo{volume}{52}},
  \bibinfo{pages}{14636} (\bibinfo{year}{1995}).

\bibitem[{\citenamefont{Sipe and Shkrebtii}(2000)}]{PhysRevB.61.5337}
\bibinfo{author}{\bibfnamefont{J.~E.} \bibnamefont{Sipe}} \bibnamefont{and}
  \bibinfo{author}{\bibfnamefont{A.~I.} \bibnamefont{Shkrebtii}},
  \bibinfo{journal}{Phys. Rev. B} \textbf{\bibinfo{volume}{61}},
  \bibinfo{pages}{5337} (\bibinfo{year}{2000}).

\bibitem[{\citenamefont{Sturman}(2020)}]{Sturman2020}
\bibinfo{author}{\bibfnamefont{B.~I.} \bibnamefont{Sturman}},
  \bibinfo{journal}{Physics-Uspekhi} \textbf{\bibinfo{volume}{63}},
  \bibinfo{pages}{407} (\bibinfo{year}{2020}).

\bibitem[{\citenamefont{Young and Rappe}(2012)}]{PhysRevLett.109.116601}
\bibinfo{author}{\bibfnamefont{S.~M.} \bibnamefont{Young}} \bibnamefont{and}
  \bibinfo{author}{\bibfnamefont{A.~M.} \bibnamefont{Rappe}},
  \bibinfo{journal}{Phys. Rev. Lett.} \textbf{\bibinfo{volume}{109}},
  \bibinfo{pages}{116601} (\bibinfo{year}{2012}).

\bibitem[{\citenamefont{Sodemann and Fu}(2015)}]{PhysRevLett.115.216806}
\bibinfo{author}{\bibfnamefont{I.}~\bibnamefont{Sodemann}} \bibnamefont{and}
  \bibinfo{author}{\bibfnamefont{L.}~\bibnamefont{Fu}}, \bibinfo{journal}{Phys.
  Rev. Lett.} \textbf{\bibinfo{volume}{115}}, \bibinfo{pages}{216806}
  (\bibinfo{year}{2015}).

\bibitem[{\citenamefont{Morimoto and Nagaosa}(2016)}]{Morimotoe1501524}
\bibinfo{author}{\bibfnamefont{T.}~\bibnamefont{Morimoto}} \bibnamefont{and}
  \bibinfo{author}{\bibfnamefont{N.}~\bibnamefont{Nagaosa}},
  \bibinfo{journal}{Science Advances} \textbf{\bibinfo{volume}{2}}
  (\bibinfo{year}{2016}).

\bibitem[{\citenamefont{Nagaosa and Morimoto}(2017)}]{NaMo}
\bibinfo{author}{\bibfnamefont{N.}~\bibnamefont{Nagaosa}} \bibnamefont{and}
  \bibinfo{author}{\bibfnamefont{T.}~\bibnamefont{Morimoto}},
  \bibinfo{journal}{Advanced Materials} \textbf{\bibinfo{volume}{29}},
  \bibinfo{pages}{1603345} (\bibinfo{year}{2017}).

\bibitem[{\citenamefont{Matsyshyn and Sodemann}(2019)}]{PhysRevLett.123.246602}
\bibinfo{author}{\bibfnamefont{O.}~\bibnamefont{Matsyshyn}} \bibnamefont{and}
  \bibinfo{author}{\bibfnamefont{I.}~\bibnamefont{Sodemann}},
  \bibinfo{journal}{Phys. Rev. Lett.} \textbf{\bibinfo{volume}{123}},
  \bibinfo{pages}{246602} (\bibinfo{year}{2019}).

\bibitem[{\citenamefont{Parker et~al.}(2019)\citenamefont{Parker, Morimoto,
  Orenstein, and Moore}}]{PhysRevB.99.045121}
\bibinfo{author}{\bibfnamefont{D.~E.} \bibnamefont{Parker}},
  \bibinfo{author}{\bibfnamefont{T.}~\bibnamefont{Morimoto}},
  \bibinfo{author}{\bibfnamefont{J.}~\bibnamefont{Orenstein}},
  \bibnamefont{and} \bibinfo{author}{\bibfnamefont{J.~E.} \bibnamefont{Moore}},
  \bibinfo{journal}{Phys. Rev. B} \textbf{\bibinfo{volume}{99}},
  \bibinfo{pages}{045121} (\bibinfo{year}{2019}).

\bibitem[{\citenamefont{de~Juan et~al.}(2017)\citenamefont{de~Juan, Grushin,
  Morimoto, and Moore}}]{deJuan2017}
\bibinfo{author}{\bibfnamefont{F.}~\bibnamefont{de~Juan}},
  \bibinfo{author}{\bibfnamefont{A.~G.} \bibnamefont{Grushin}},
  \bibinfo{author}{\bibfnamefont{T.}~\bibnamefont{Morimoto}}, \bibnamefont{and}
  \bibinfo{author}{\bibfnamefont{J.~E.} \bibnamefont{Moore}},
  \bibinfo{journal}{Nature Communications} \textbf{\bibinfo{volume}{8}},
  \bibinfo{pages}{15995 EP } (\bibinfo{year}{2017}), \bibinfo{note}{article}.

\bibitem[{\citenamefont{Matsyshyn et~al.}(2021)\citenamefont{Matsyshyn, Dey,
  Sodemann, and Sun}}]{matsyshyn2020berry}
\bibinfo{author}{\bibfnamefont{O.}~\bibnamefont{Matsyshyn}},
  \bibinfo{author}{\bibfnamefont{U.}~\bibnamefont{Dey}},
  \bibinfo{author}{\bibfnamefont{I.}~\bibnamefont{Sodemann}}, \bibnamefont{and}
  \bibinfo{author}{\bibfnamefont{Y.}~\bibnamefont{Sun}}, \bibinfo{journal}{J.
  Phys. D} \textbf{\bibinfo{volume}{54}}, \bibinfo{pages}{404001}
  (\bibinfo{year}{2021}).

\bibitem[{\citenamefont{Chan et~al.}(2017)\citenamefont{Chan, Lindner, Refael,
  and Lee}}]{PhysRevB.95.041104}
\bibinfo{author}{\bibfnamefont{C.}~\bibnamefont{Chan}},
  \bibinfo{author}{\bibfnamefont{N.~H.} \bibnamefont{Lindner}},
  \bibinfo{author}{\bibfnamefont{G.}~\bibnamefont{Refael}}, \bibnamefont{and}
  \bibinfo{author}{\bibfnamefont{P.~A.} \bibnamefont{Lee}},
  \bibinfo{journal}{Phys. Rev. B} \textbf{\bibinfo{volume}{95}},
  \bibinfo{pages}{041104} (\bibinfo{year}{2017}).

\bibitem[{\citenamefont{Vanderbilt}(2018)}]{vanderbilt2018berry}
\bibinfo{author}{\bibfnamefont{D.}~\bibnamefont{Vanderbilt}},
  \emph{\bibinfo{title}{Berry Phases in Electronic Structure Theory: Electric
  Polarization, Orbital Magnetization and Topological Insulators}}, Titolo
  collana (\bibinfo{publisher}{Cambridge University Press},
  \bibinfo{year}{2018}).

\bibitem[{\citenamefont{Moore and Orenstein}(2010)}]{PhysRevLett.105.026805}
\bibinfo{author}{\bibfnamefont{J.~E.} \bibnamefont{Moore}} \bibnamefont{and}
  \bibinfo{author}{\bibfnamefont{J.}~\bibnamefont{Orenstein}},
  \bibinfo{journal}{Phys. Rev. Lett.} \textbf{\bibinfo{volume}{105}},
  \bibinfo{pages}{026805} (\bibinfo{year}{2010}).

\bibitem[{\citenamefont{Kang et~al.}(2019)\citenamefont{Kang, Li, Sohn, Shan,
  and Mak}}]{Kang2019}
\bibinfo{author}{\bibfnamefont{K.}~\bibnamefont{Kang}},
  \bibinfo{author}{\bibfnamefont{T.}~\bibnamefont{Li}},
  \bibinfo{author}{\bibfnamefont{E.}~\bibnamefont{Sohn}},
  \bibinfo{author}{\bibfnamefont{J.}~\bibnamefont{Shan}}, \bibnamefont{and}
  \bibinfo{author}{\bibfnamefont{K.~F.} \bibnamefont{Mak}},
  \bibinfo{journal}{Nature Materials} \textbf{\bibinfo{volume}{18}},
  \bibinfo{pages}{324} (\bibinfo{year}{2019}), ISSN \bibinfo{issn}{1476-4660}.

\bibitem[{\citenamefont{Ma et~al.}(2019)\citenamefont{Ma, Xu, Shen, MacNeill,
  Fatemi, Chang, Mier~Valdivia, Wu, Du, Hsu et~al.}}]{Ma2019}
\bibinfo{author}{\bibfnamefont{Q.}~\bibnamefont{Ma}},
  \bibinfo{author}{\bibfnamefont{S.}~\bibnamefont{Xu}},
  \bibinfo{author}{\bibfnamefont{H.}~\bibnamefont{Shen}},
  \bibinfo{author}{\bibfnamefont{D.}~\bibnamefont{MacNeill}},
  \bibinfo{author}{\bibfnamefont{V.}~\bibnamefont{Fatemi}},
  \bibinfo{author}{\bibfnamefont{T.}~\bibnamefont{Chang}},
  \bibinfo{author}{\bibfnamefont{A.~M.} \bibnamefont{Mier~Valdivia}},
  \bibinfo{author}{\bibfnamefont{S.}~\bibnamefont{Wu}},
  \bibinfo{author}{\bibfnamefont{Z.}~\bibnamefont{Du}},
  \bibinfo{author}{\bibfnamefont{C.}~\bibnamefont{Hsu}}, \bibnamefont{et~al.},
  \bibinfo{journal}{Nature} \textbf{\bibinfo{volume}{565}},
  \bibinfo{pages}{337} (\bibinfo{year}{2019}), ISSN \bibinfo{issn}{1476-4687}.

\bibitem[{\citenamefont{Brehm et~al.}(2014)\citenamefont{Brehm, Young, Zheng,
  and Rappe}}]{BrehmYoung}
\bibinfo{author}{\bibfnamefont{J.~A.} \bibnamefont{Brehm}},
  \bibinfo{author}{\bibfnamefont{S.~M.} \bibnamefont{Young}},
  \bibinfo{author}{\bibfnamefont{F.}~\bibnamefont{Zheng}}, \bibnamefont{and}
  \bibinfo{author}{\bibfnamefont{A.~M.} \bibnamefont{Rappe}},
  \bibinfo{journal}{The Journal of Chemical Physics}
  \textbf{\bibinfo{volume}{141}}, \bibinfo{pages}{204704}
  (\bibinfo{year}{2014}).

\bibitem[{\citenamefont{Rangel et~al.}(2017)\citenamefont{Rangel, Fregoso,
  Mendoza, Morimoto, Moore, and Neaton}}]{PhysRevLett.119.067402}
\bibinfo{author}{\bibfnamefont{T.}~\bibnamefont{Rangel}},
  \bibinfo{author}{\bibfnamefont{B.~M.} \bibnamefont{Fregoso}},
  \bibinfo{author}{\bibfnamefont{B.~S.} \bibnamefont{Mendoza}},
  \bibinfo{author}{\bibfnamefont{T.}~\bibnamefont{Morimoto}},
  \bibinfo{author}{\bibfnamefont{J.~E.} \bibnamefont{Moore}}, \bibnamefont{and}
  \bibinfo{author}{\bibfnamefont{J.~B.} \bibnamefont{Neaton}},
  \bibinfo{journal}{Phys. Rev. Lett.} \textbf{\bibinfo{volume}{119}},
  \bibinfo{pages}{067402} (\bibinfo{year}{2017}).

\bibitem[{\citenamefont{Cook et~al.}(2017)\citenamefont{Cook, M.~Fregoso,
  de~Juan, Coh, and Moore}}]{Cook2017}
\bibinfo{author}{\bibfnamefont{A.~M.} \bibnamefont{Cook}},
  \bibinfo{author}{\bibfnamefont{B.}~\bibnamefont{M.~Fregoso}},
  \bibinfo{author}{\bibfnamefont{F.}~\bibnamefont{de~Juan}},
  \bibinfo{author}{\bibfnamefont{S.}~\bibnamefont{Coh}}, \bibnamefont{and}
  \bibinfo{author}{\bibfnamefont{J.~E.} \bibnamefont{Moore}},
  \bibinfo{journal}{Nature Communications} \textbf{\bibinfo{volume}{8}},
  \bibinfo{pages}{14176 EP } (\bibinfo{year}{2017}), \bibinfo{note}{article}.

\bibitem[{\citenamefont{Morimoto et~al.}(2018)\citenamefont{Morimoto, Nakamura,
  Kawasaki, and Nagaosa}}]{PhysRevLett.121.267401}
\bibinfo{author}{\bibfnamefont{T.}~\bibnamefont{Morimoto}},
  \bibinfo{author}{\bibfnamefont{M.}~\bibnamefont{Nakamura}},
  \bibinfo{author}{\bibfnamefont{M.}~\bibnamefont{Kawasaki}}, \bibnamefont{and}
  \bibinfo{author}{\bibfnamefont{N.}~\bibnamefont{Nagaosa}},
  \bibinfo{journal}{Phys. Rev. Lett.} \textbf{\bibinfo{volume}{121}},
  \bibinfo{pages}{267401} (\bibinfo{year}{2018}).

\bibitem[{\citenamefont{Kumar et~al.}(2021)\citenamefont{Kumar, Hsu, Sharma,
  Chang, Yu, Wang, Eda, Liang, and Yang}}]{Kumar2021}
\bibinfo{author}{\bibfnamefont{D.}~\bibnamefont{Kumar}},
  \bibinfo{author}{\bibfnamefont{C.-H.} \bibnamefont{Hsu}},
  \bibinfo{author}{\bibfnamefont{R.}~\bibnamefont{Sharma}},
  \bibinfo{author}{\bibfnamefont{T.-R.} \bibnamefont{Chang}},
  \bibinfo{author}{\bibfnamefont{P.}~\bibnamefont{Yu}},
  \bibinfo{author}{\bibfnamefont{J.}~\bibnamefont{Wang}},
  \bibinfo{author}{\bibfnamefont{G.}~\bibnamefont{Eda}},
  \bibinfo{author}{\bibfnamefont{G.}~\bibnamefont{Liang}}, \bibnamefont{and}
  \bibinfo{author}{\bibfnamefont{H.}~\bibnamefont{Yang}},
  \bibinfo{journal}{Nature Nanotechnology} \textbf{\bibinfo{volume}{16}},
  \bibinfo{pages}{421–425} (\bibinfo{year}{2021}), ISSN
  \bibinfo{issn}{1748-3395}.

\bibitem[{\citenamefont{Kitamura et~al.}(2020)\citenamefont{Kitamura, Nagaosa,
  and Morimoto}}]{2020arXiv200903596K}
\bibinfo{author}{\bibfnamefont{S.}~\bibnamefont{Kitamura}},
  \bibinfo{author}{\bibfnamefont{N.}~\bibnamefont{Nagaosa}}, \bibnamefont{and}
  \bibinfo{author}{\bibfnamefont{T.}~\bibnamefont{Morimoto}},
  \bibinfo{journal}{Phys. Rev. B} \textbf{\bibinfo{volume}{102}},
  \bibinfo{pages}{245141} (\bibinfo{year}{2020}).

\bibitem[{\citenamefont{Afonin et~al.}(1995)\citenamefont{Afonin, Gurevich, and
  Laiho}}]{PhysRevB.52.2090}
\bibinfo{author}{\bibfnamefont{V.~V.} \bibnamefont{Afonin}},
  \bibinfo{author}{\bibfnamefont{V.~L.} \bibnamefont{Gurevich}},
  \bibnamefont{and} \bibinfo{author}{\bibfnamefont{R.}~\bibnamefont{Laiho}},
  \bibinfo{journal}{Phys. Rev. B} \textbf{\bibinfo{volume}{52}},
  \bibinfo{pages}{2090} (\bibinfo{year}{1995}).

\bibitem[{\citenamefont{Leppenen et~al.}(2019)\citenamefont{Leppenen, Ivchenko,
  and Golub}}]{Leppenen2019}
\bibinfo{author}{\bibfnamefont{N.~V.} \bibnamefont{Leppenen}},
  \bibinfo{author}{\bibfnamefont{E.~L.} \bibnamefont{Ivchenko}},
  \bibnamefont{and} \bibinfo{author}{\bibfnamefont{L.~E.} \bibnamefont{Golub}},
  \bibinfo{journal}{physica status solidi (b)} \textbf{\bibinfo{volume}{256}},
  \bibinfo{pages}{1900305} (\bibinfo{year}{2019}).

\bibitem[{\citenamefont{Dantas et~al.}(2021)\citenamefont{Dantas, Wang,
  Sur{\'o}wka, and Oka}}]{dantas2020nonperturbative}
\bibinfo{author}{\bibfnamefont{R.~M.~A.} \bibnamefont{Dantas}},
  \bibinfo{author}{\bibfnamefont{Z.}~\bibnamefont{Wang}},
  \bibinfo{author}{\bibfnamefont{P.}~\bibnamefont{Sur{\'o}wka}},
  \bibnamefont{and} \bibinfo{author}{\bibfnamefont{T.}~\bibnamefont{Oka}},
  \emph{\bibinfo{title}{Nonperturbative topological current in weyl and dirac
  semimetals in laser fields}} (\bibinfo{year}{2021}).

\bibitem[{\citenamefont{James and Smith}(1979)}]{PhysRevLett.42.1495}
\bibinfo{author}{\bibfnamefont{R.~B.} \bibnamefont{James}} \bibnamefont{and}
  \bibinfo{author}{\bibfnamefont{D.~L.} \bibnamefont{Smith}},
  \bibinfo{journal}{Phys. Rev. Lett.} \textbf{\bibinfo{volume}{42}},
  \bibinfo{pages}{1495} (\bibinfo{year}{1979}).

\bibitem[{\citenamefont{James and Smith}(1980)}]{PhysRevB.21.3502}
\bibinfo{author}{\bibfnamefont{R.~B.} \bibnamefont{James}} \bibnamefont{and}
  \bibinfo{author}{\bibfnamefont{D.~L.} \bibnamefont{Smith}},
  \bibinfo{journal}{Phys. Rev. B} \textbf{\bibinfo{volume}{21}},
  \bibinfo{pages}{3502} (\bibinfo{year}{1980}).

\bibitem[{\citenamefont{Parshin and Shabaev}(1987)}]{Parshin}
\bibinfo{author}{\bibfnamefont{D.}~\bibnamefont{Parshin}} \bibnamefont{and}
  \bibinfo{author}{\bibfnamefont{A.}~\bibnamefont{Shabaev}},
  \bibinfo{journal}{Zh. Eksp. Teor. Fiz.} \textbf{\bibinfo{volume}{92}},
  \bibinfo{pages}{1471} (\bibinfo{year}{1987}).

\bibitem[{\citenamefont{Gerchikov et~al.}(1989)\citenamefont{Gerchikov,
  Parshin, and Shabayev}}]{Parshin2}
\bibinfo{author}{\bibfnamefont{L.}~\bibnamefont{Gerchikov}},
  \bibinfo{author}{\bibfnamefont{D.}~\bibnamefont{Parshin}}, \bibnamefont{and}
  \bibinfo{author}{\bibfnamefont{A.}~\bibnamefont{Shabayev}},
  \bibinfo{journal}{Zh. Eksp. Teor. Fiz} \textbf{\bibinfo{volume}{96}},
  \bibinfo{pages}{1046} (\bibinfo{year}{1989}).

\bibitem[{\citenamefont{Oka and Aoki}(2009)}]{KelFloq6}
\bibinfo{author}{\bibfnamefont{T.}~\bibnamefont{Oka}} \bibnamefont{and}
  \bibinfo{author}{\bibfnamefont{H.}~\bibnamefont{Aoki}},
  \bibinfo{journal}{Phys. Rev. B} \textbf{\bibinfo{volume}{79}},
  \bibinfo{pages}{081406} (\bibinfo{year}{2009}).

\bibitem[{\citenamefont{Jauho et~al.}(1994{\natexlab{a}})\citenamefont{Jauho,
  Wingreen, and Meir}}]{PhysRevB.50.5528}
\bibinfo{author}{\bibfnamefont{A.-P.} \bibnamefont{Jauho}},
  \bibinfo{author}{\bibfnamefont{N.~S.} \bibnamefont{Wingreen}},
  \bibnamefont{and} \bibinfo{author}{\bibfnamefont{Y.}~\bibnamefont{Meir}},
  \bibinfo{journal}{Phys. Rev. B} \textbf{\bibinfo{volume}{50}},
  \bibinfo{pages}{5528} (\bibinfo{year}{1994}{\natexlab{a}}).

\bibitem[{\citenamefont{Kamenev}(2011)}]{kamenev_2011}
\bibinfo{author}{\bibfnamefont{A.}~\bibnamefont{Kamenev}},
  \emph{\bibinfo{title}{Field Theory of Non-Equilibrium Systems}}
  (\bibinfo{publisher}{Cambridge University Press}, \bibinfo{year}{2011}).

\bibitem[{\citenamefont{Lazarides
  et~al.}(2014{\natexlab{a}})\citenamefont{Lazarides, Das, and
  Moessner}}]{Moess4}
\bibinfo{author}{\bibfnamefont{A.}~\bibnamefont{Lazarides}},
  \bibinfo{author}{\bibfnamefont{A.}~\bibnamefont{Das}}, \bibnamefont{and}
  \bibinfo{author}{\bibfnamefont{R.}~\bibnamefont{Moessner}},
  \bibinfo{journal}{Phys. Rev. Lett.} \textbf{\bibinfo{volume}{112}},
  \bibinfo{pages}{150401} (\bibinfo{year}{2014}{\natexlab{a}}).

\bibitem[{\citenamefont{Russomanno et~al.}(2012)\citenamefont{Russomanno,
  Silva, and Santoro}}]{PhysRevLett.109.257201}
\bibinfo{author}{\bibfnamefont{A.}~\bibnamefont{Russomanno}},
  \bibinfo{author}{\bibfnamefont{A.}~\bibnamefont{Silva}}, \bibnamefont{and}
  \bibinfo{author}{\bibfnamefont{G.~E.} \bibnamefont{Santoro}},
  \bibinfo{journal}{Phys. Rev. Lett.} \textbf{\bibinfo{volume}{109}},
  \bibinfo{pages}{257201} (\bibinfo{year}{2012}).

\bibitem[{\citenamefont{Fregoso et~al.}(2013)\citenamefont{Fregoso, Wang,
  Gedik, and Galitski}}]{PhysRevB.88.155129}
\bibinfo{author}{\bibfnamefont{B.~M.} \bibnamefont{Fregoso}},
  \bibinfo{author}{\bibfnamefont{Y.~H.} \bibnamefont{Wang}},
  \bibinfo{author}{\bibfnamefont{N.}~\bibnamefont{Gedik}}, \bibnamefont{and}
  \bibinfo{author}{\bibfnamefont{V.}~\bibnamefont{Galitski}},
  \bibinfo{journal}{Phys. Rev. B} \textbf{\bibinfo{volume}{88}},
  \bibinfo{pages}{155129} (\bibinfo{year}{2013}).

\bibitem[{\citenamefont{{Kamenev}}(2004)}]{KelFloq1}
\bibinfo{author}{\bibfnamefont{A.}~\bibnamefont{{Kamenev}}},
  \bibinfo{journal}{arXiv e-prints} pp. \bibinfo{pages}{cond--mat/0412296}
  (\bibinfo{year}{2004}).

\bibitem[{\citenamefont{Johnsen and Jauho}(1999)}]{KelFloq2}
\bibinfo{author}{\bibfnamefont{K.}~\bibnamefont{Johnsen}} \bibnamefont{and}
  \bibinfo{author}{\bibfnamefont{A.-P.} \bibnamefont{Jauho}},
  \bibinfo{journal}{Phys. Rev. Lett.} \textbf{\bibinfo{volume}{83}},
  \bibinfo{pages}{1207} (\bibinfo{year}{1999}).

\bibitem[{\citenamefont{Jauho et~al.}(1994{\natexlab{b}})\citenamefont{Jauho,
  Wingreen, and Meir}}]{KelFloq3}
\bibinfo{author}{\bibfnamefont{A.-P.} \bibnamefont{Jauho}},
  \bibinfo{author}{\bibfnamefont{N.~S.} \bibnamefont{Wingreen}},
  \bibnamefont{and} \bibinfo{author}{\bibfnamefont{Y.}~\bibnamefont{Meir}},
  \bibinfo{journal}{Phys. Rev. B} \textbf{\bibinfo{volume}{50}},
  \bibinfo{pages}{5528} (\bibinfo{year}{1994}{\natexlab{b}}).

\bibitem[{\citenamefont{Kohler et~al.}(2005)\citenamefont{Kohler, Lehmann, and
  Hanggi}}]{KelFloq4}
\bibinfo{author}{\bibfnamefont{S.}~\bibnamefont{Kohler}},
  \bibinfo{author}{\bibfnamefont{J.}~\bibnamefont{Lehmann}}, \bibnamefont{and}
  \bibinfo{author}{\bibfnamefont{P.}~\bibnamefont{Hanggi}},
  \bibinfo{journal}{Physics Reports} \textbf{\bibinfo{volume}{406}},
  \bibinfo{pages}{379} (\bibinfo{year}{2005}).

\bibitem[{\citenamefont{Kitagawa et~al.}(2011)\citenamefont{Kitagawa, Oka,
  Brataas, Fu, and Demler}}]{KelFloq5}
\bibinfo{author}{\bibfnamefont{T.}~\bibnamefont{Kitagawa}},
  \bibinfo{author}{\bibfnamefont{T.}~\bibnamefont{Oka}},
  \bibinfo{author}{\bibfnamefont{A.}~\bibnamefont{Brataas}},
  \bibinfo{author}{\bibfnamefont{L.}~\bibnamefont{Fu}}, \bibnamefont{and}
  \bibinfo{author}{\bibfnamefont{E.}~\bibnamefont{Demler}},
  \bibinfo{journal}{Phys. Rev. B} \textbf{\bibinfo{volume}{84}},
  \bibinfo{pages}{235108} (\bibinfo{year}{2011}).

\bibitem[{\citenamefont{Khemani et~al.}(2016)\citenamefont{Khemani, Lazarides,
  Moessner, and Sondhi}}]{Moess1}
\bibinfo{author}{\bibfnamefont{V.}~\bibnamefont{Khemani}},
  \bibinfo{author}{\bibfnamefont{A.}~\bibnamefont{Lazarides}},
  \bibinfo{author}{\bibfnamefont{R.}~\bibnamefont{Moessner}}, \bibnamefont{and}
  \bibinfo{author}{\bibfnamefont{S.~L.} \bibnamefont{Sondhi}},
  \bibinfo{journal}{Phys. Rev. Lett.} \textbf{\bibinfo{volume}{116}},
  \bibinfo{pages}{250401} (\bibinfo{year}{2016}).

\bibitem[{\citenamefont{Lazarides
  et~al.}(2014{\natexlab{b}})\citenamefont{Lazarides, Das, and
  Moessner}}]{Moess2}
\bibinfo{author}{\bibfnamefont{A.}~\bibnamefont{Lazarides}},
  \bibinfo{author}{\bibfnamefont{A.}~\bibnamefont{Das}}, \bibnamefont{and}
  \bibinfo{author}{\bibfnamefont{R.}~\bibnamefont{Moessner}},
  \bibinfo{journal}{Phys. Rev. E} \textbf{\bibinfo{volume}{90}},
  \bibinfo{pages}{012110} (\bibinfo{year}{2014}{\natexlab{b}}).

\bibitem[{\citenamefont{Lazarides et~al.}(2015)\citenamefont{Lazarides, Das,
  and Moessner}}]{Moess3}
\bibinfo{author}{\bibfnamefont{A.}~\bibnamefont{Lazarides}},
  \bibinfo{author}{\bibfnamefont{A.}~\bibnamefont{Das}}, \bibnamefont{and}
  \bibinfo{author}{\bibfnamefont{R.}~\bibnamefont{Moessner}},
  \bibinfo{journal}{Phys. Rev. Lett.} \textbf{\bibinfo{volume}{115}},
  \bibinfo{pages}{030402} (\bibinfo{year}{2015}).

\bibitem[{\citenamefont{Ahn et~al.}(2020)\citenamefont{Ahn, Guo, and
  Nagaosa}}]{ahn2020lowfrequency}
\bibinfo{author}{\bibfnamefont{J.}~\bibnamefont{Ahn}},
  \bibinfo{author}{\bibfnamefont{G.-Y.} \bibnamefont{Guo}}, \bibnamefont{and}
  \bibinfo{author}{\bibfnamefont{N.}~\bibnamefont{Nagaosa}},
  \bibinfo{journal}{Phys. Rev. X} \textbf{\bibinfo{volume}{10}},
  \bibinfo{pages}{041041} (\bibinfo{year}{2020}).

\bibitem[{\citenamefont{Hornung and von Baltz}(2021)}]{hornung2020quantum}
\bibinfo{author}{\bibfnamefont{D.}~\bibnamefont{Hornung}} \bibnamefont{and}
  \bibinfo{author}{\bibfnamefont{R.}~\bibnamefont{von Baltz}},
  \emph{\bibinfo{title}{Quantum kinetics of the magnetophotogalvanic effect}}
  (\bibinfo{year}{2021}).

\bibitem[{\citenamefont{Ivchenko and Pikus}(1978)}]{1978JETPL..27..604I}
\bibinfo{author}{\bibfnamefont{E.}~\bibnamefont{Ivchenko}} \bibnamefont{and}
  \bibinfo{author}{\bibfnamefont{G.}~\bibnamefont{Pikus}},
  \bibinfo{journal}{Soviet Journal of Experimental and Theoretical Physics
  Letters} \textbf{\bibinfo{volume}{27}}, \bibinfo{pages}{604}
  (\bibinfo{year}{1978}).

\bibitem[{\citenamefont{Zhang et~al.}(2019)\citenamefont{Zhang, Holder,
  Ishizuka, de~Juan, Nagaosa, Felser, and Yan}}]{Zhang2019}
\bibinfo{author}{\bibfnamefont{Y.}~\bibnamefont{Zhang}},
  \bibinfo{author}{\bibfnamefont{T.}~\bibnamefont{Holder}},
  \bibinfo{author}{\bibfnamefont{H.}~\bibnamefont{Ishizuka}},
  \bibinfo{author}{\bibfnamefont{F.}~\bibnamefont{de~Juan}},
  \bibinfo{author}{\bibfnamefont{N.}~\bibnamefont{Nagaosa}},
  \bibinfo{author}{\bibfnamefont{C.}~\bibnamefont{Felser}}, \bibnamefont{and}
  \bibinfo{author}{\bibfnamefont{B.}~\bibnamefont{Yan}},
  \bibinfo{journal}{Nature Communications} \textbf{\bibinfo{volume}{10}},
  \bibinfo{pages}{3783} (\bibinfo{year}{2019}), ISSN \bibinfo{issn}{2041-1723}.

\bibitem[{\citenamefont{Golub and Ivchenko}(2018)}]{PhysRevB.98.075305}
\bibinfo{author}{\bibfnamefont{L.~E.} \bibnamefont{Golub}} \bibnamefont{and}
  \bibinfo{author}{\bibfnamefont{E.~L.} \bibnamefont{Ivchenko}},
  \bibinfo{journal}{Phys. Rev. B} \textbf{\bibinfo{volume}{98}},
  \bibinfo{pages}{075305} (\bibinfo{year}{2018}).

\bibitem[{\citenamefont{Flicker et~al.}(2018)\citenamefont{Flicker, de~Juan,
  Bradlyn, Morimoto, Vergniory, and Grushin}}]{PhysRevB.98.155145}
\bibinfo{author}{\bibfnamefont{F.}~\bibnamefont{Flicker}},
  \bibinfo{author}{\bibfnamefont{F.}~\bibnamefont{de~Juan}},
  \bibinfo{author}{\bibfnamefont{B.}~\bibnamefont{Bradlyn}},
  \bibinfo{author}{\bibfnamefont{T.}~\bibnamefont{Morimoto}},
  \bibinfo{author}{\bibfnamefont{M.~G.} \bibnamefont{Vergniory}},
  \bibnamefont{and} \bibinfo{author}{\bibfnamefont{A.~G.}
  \bibnamefont{Grushin}}, \bibinfo{journal}{Phys. Rev. B}
  \textbf{\bibinfo{volume}{98}}, \bibinfo{pages}{155145}
  (\bibinfo{year}{2018}).

\bibitem[{\citenamefont{de~Juan et~al.}(2020)\citenamefont{de~Juan, Zhang,
  Morimoto, Sun, Moore, and Grushin}}]{conv}
\bibinfo{author}{\bibfnamefont{F.}~\bibnamefont{de~Juan}},
  \bibinfo{author}{\bibfnamefont{Y.}~\bibnamefont{Zhang}},
  \bibinfo{author}{\bibfnamefont{T.}~\bibnamefont{Morimoto}},
  \bibinfo{author}{\bibfnamefont{Y.}~\bibnamefont{Sun}},
  \bibinfo{author}{\bibfnamefont{J.~E.} \bibnamefont{Moore}}, \bibnamefont{and}
  \bibinfo{author}{\bibfnamefont{A.~G.} \bibnamefont{Grushin}},
  \bibinfo{journal}{Phys. Rev. Research} \textbf{\bibinfo{volume}{2}},
  \bibinfo{pages}{012017} (\bibinfo{year}{2020}).

\bibitem[{\citenamefont{Avdoshkin et~al.}(2020)\citenamefont{Avdoshkin, Kozii,
  and Moore}}]{PhysRevLett.124.196603}
\bibinfo{author}{\bibfnamefont{A.}~\bibnamefont{Avdoshkin}},
  \bibinfo{author}{\bibfnamefont{V.}~\bibnamefont{Kozii}}, \bibnamefont{and}
  \bibinfo{author}{\bibfnamefont{J.~E.} \bibnamefont{Moore}},
  \bibinfo{journal}{Phys. Rev. Lett.} \textbf{\bibinfo{volume}{124}},
  \bibinfo{pages}{196603} (\bibinfo{year}{2020}).

\bibitem[{\citenamefont{Rees et~al.}(2020)\citenamefont{Rees, Manna, Lu,
  Morimoto, Borrmann, Felser, Moore, Torchinsky, and Orenstein}}]{Rees2020}
\bibinfo{author}{\bibfnamefont{D.}~\bibnamefont{Rees}},
  \bibinfo{author}{\bibfnamefont{K.}~\bibnamefont{Manna}},
  \bibinfo{author}{\bibfnamefont{B.}~\bibnamefont{Lu}},
  \bibinfo{author}{\bibfnamefont{T.}~\bibnamefont{Morimoto}},
  \bibinfo{author}{\bibfnamefont{H.}~\bibnamefont{Borrmann}},
  \bibinfo{author}{\bibfnamefont{C.}~\bibnamefont{Felser}},
  \bibinfo{author}{\bibfnamefont{J.~E.} \bibnamefont{Moore}},
  \bibinfo{author}{\bibfnamefont{D.~H.} \bibnamefont{Torchinsky}},
  \bibnamefont{and}
  \bibinfo{author}{\bibfnamefont{J.}~\bibnamefont{Orenstein}},
  \bibinfo{journal}{Science Advances} \textbf{\bibinfo{volume}{6}},
  \bibinfo{pages}{eaba0509} (\bibinfo{year}{2020}).

\bibitem[{\citenamefont{Ni et~al.}(2021{\natexlab{a}})\citenamefont{Ni, Wang,
  Zhang, Pozo, Xu, Han, Manna, Paglione, Felser, Grushin et~al.}}]{ni2020giant}
\bibinfo{author}{\bibfnamefont{Z.}~\bibnamefont{Ni}},
  \bibinfo{author}{\bibfnamefont{K.}~\bibnamefont{Wang}},
  \bibinfo{author}{\bibfnamefont{Y.}~\bibnamefont{Zhang}},
  \bibinfo{author}{\bibfnamefont{O.}~\bibnamefont{Pozo}},
  \bibinfo{author}{\bibfnamefont{B.}~\bibnamefont{Xu}},
  \bibinfo{author}{\bibfnamefont{X.}~\bibnamefont{Han}},
  \bibinfo{author}{\bibfnamefont{K.}~\bibnamefont{Manna}},
  \bibinfo{author}{\bibfnamefont{J.}~\bibnamefont{Paglione}},
  \bibinfo{author}{\bibfnamefont{C.}~\bibnamefont{Felser}},
  \bibinfo{author}{\bibfnamefont{A.~G.} \bibnamefont{Grushin}},
  \bibnamefont{et~al.}, \bibinfo{journal}{Nature Communications}
  \textbf{\bibinfo{volume}{12}}, \bibinfo{pages}{154}
  (\bibinfo{year}{2021}{\natexlab{a}}), ISSN \bibinfo{issn}{2041-1723}.

\bibitem[{\citenamefont{Ni et~al.}(2021{\natexlab{b}})\citenamefont{Ni, Wang,
  Zhang, Pozo, Xu, Han, Manna, Paglione, Felser, Grushin et~al.}}]{Ni2021}
\bibinfo{author}{\bibfnamefont{Z.}~\bibnamefont{Ni}},
  \bibinfo{author}{\bibfnamefont{K.}~\bibnamefont{Wang}},
  \bibinfo{author}{\bibfnamefont{Y.}~\bibnamefont{Zhang}},
  \bibinfo{author}{\bibfnamefont{O.}~\bibnamefont{Pozo}},
  \bibinfo{author}{\bibfnamefont{B.}~\bibnamefont{Xu}},
  \bibinfo{author}{\bibfnamefont{X.}~\bibnamefont{Han}},
  \bibinfo{author}{\bibfnamefont{K.}~\bibnamefont{Manna}},
  \bibinfo{author}{\bibfnamefont{J.}~\bibnamefont{Paglione}},
  \bibinfo{author}{\bibfnamefont{C.}~\bibnamefont{Felser}},
  \bibinfo{author}{\bibfnamefont{A.~G.} \bibnamefont{Grushin}},
  \bibnamefont{et~al.}, \bibinfo{journal}{Nature Communications}
  \textbf{\bibinfo{volume}{12}}, \bibinfo{pages}{154}
  (\bibinfo{year}{2021}{\natexlab{b}}), ISSN \bibinfo{issn}{2041-1723}.

\bibitem[{\citenamefont{Ma et~al.}(2017)\citenamefont{Ma, Xu, Chan, Zhang,
  Chang, Lin, Xie, Palacios, Lin, Jia et~al.}}]{Ma2017}
\bibinfo{author}{\bibfnamefont{Q.}~\bibnamefont{Ma}},
  \bibinfo{author}{\bibfnamefont{S.}~\bibnamefont{Xu}},
  \bibinfo{author}{\bibfnamefont{C.}~\bibnamefont{Chan}},
  \bibinfo{author}{\bibfnamefont{C.}~\bibnamefont{Zhang}},
  \bibinfo{author}{\bibfnamefont{G.}~\bibnamefont{Chang}},
  \bibinfo{author}{\bibfnamefont{Y.}~\bibnamefont{Lin}},
  \bibinfo{author}{\bibfnamefont{W.}~\bibnamefont{Xie}},
  \bibinfo{author}{\bibfnamefont{T.}~\bibnamefont{Palacios}},
  \bibinfo{author}{\bibfnamefont{H.}~\bibnamefont{Lin}},
  \bibinfo{author}{\bibfnamefont{S.}~\bibnamefont{Jia}}, \bibnamefont{et~al.},
  \bibinfo{journal}{Nature Physics} \textbf{\bibinfo{volume}{13}},
  \bibinfo{pages}{842 EP } (\bibinfo{year}{2017}).

\bibitem[{\citenamefont{Nagaosa et~al.}(2020)\citenamefont{Nagaosa, Morimoto,
  and Tokura}}]{Nagaosa2020}
\bibinfo{author}{\bibfnamefont{N.}~\bibnamefont{Nagaosa}},
  \bibinfo{author}{\bibfnamefont{T.}~\bibnamefont{Morimoto}}, \bibnamefont{and}
  \bibinfo{author}{\bibfnamefont{Y.}~\bibnamefont{Tokura}},
  \bibinfo{journal}{Nature Reviews Materials} \textbf{\bibinfo{volume}{5}},
  \bibinfo{pages}{621} (\bibinfo{year}{2020}), ISSN \bibinfo{issn}{2058-8437}.

\bibitem[{\citenamefont{Sun et~al.}(2017)\citenamefont{Sun, Sun, Wei, Guo,
  Tian, Chen, Yang, and Li}}]{KaiSun117203}
\bibinfo{author}{\bibfnamefont{K.}~\bibnamefont{Sun}},
  \bibinfo{author}{\bibfnamefont{S.-S.} \bibnamefont{Sun}},
  \bibinfo{author}{\bibfnamefont{L.-L.} \bibnamefont{Wei}},
  \bibinfo{author}{\bibfnamefont{C.}~\bibnamefont{Guo}},
  \bibinfo{author}{\bibfnamefont{H.-F.} \bibnamefont{Tian}},
  \bibinfo{author}{\bibfnamefont{G.-F.} \bibnamefont{Chen}},
  \bibinfo{author}{\bibfnamefont{H.-X.} \bibnamefont{Yang}}, \bibnamefont{and}
  \bibinfo{author}{\bibfnamefont{J.-Q.} \bibnamefont{Li}},
  \bibinfo{journal}{Chinese Physics Letters} \textbf{\bibinfo{volume}{34}},
  \bibinfo{eid}{117203} (\bibinfo{year}{2017}).

\bibitem[{\citenamefont{Ji et~al.}(2019)\citenamefont{Ji, Liu, Addison, Liu,
  Yu, Gao, Liu, Rappe, Kane, Mele et~al.}}]{Ji2019}
\bibinfo{author}{\bibfnamefont{Z.}~\bibnamefont{Ji}},
  \bibinfo{author}{\bibfnamefont{G.}~\bibnamefont{Liu}},
  \bibinfo{author}{\bibfnamefont{Z.}~\bibnamefont{Addison}},
  \bibinfo{author}{\bibfnamefont{W.}~\bibnamefont{Liu}},
  \bibinfo{author}{\bibfnamefont{P.}~\bibnamefont{Yu}},
  \bibinfo{author}{\bibfnamefont{H.}~\bibnamefont{Gao}},
  \bibinfo{author}{\bibfnamefont{Z.}~\bibnamefont{Liu}},
  \bibinfo{author}{\bibfnamefont{A.~M.} \bibnamefont{Rappe}},
  \bibinfo{author}{\bibfnamefont{C.~L.} \bibnamefont{Kane}},
  \bibinfo{author}{\bibfnamefont{E.~J.} \bibnamefont{Mele}},
  \bibnamefont{et~al.}, \bibinfo{journal}{Nature Materials}
  \textbf{\bibinfo{volume}{18}}, \bibinfo{pages}{955–962}
  (\bibinfo{year}{2019}), ISSN \bibinfo{issn}{1476-4660}.

\bibitem[{\citenamefont{Chang et~al.}(2017)\citenamefont{Chang, Xu, Wieder,
  Sanchez, Huang, Belopolski, Chang, Zhang, Bansil, Lin
  et~al.}}]{PhysRevLett.119.206401}
\bibinfo{author}{\bibfnamefont{G.}~\bibnamefont{Chang}},
  \bibinfo{author}{\bibfnamefont{S.-Y.} \bibnamefont{Xu}},
  \bibinfo{author}{\bibfnamefont{B.~J.} \bibnamefont{Wieder}},
  \bibinfo{author}{\bibfnamefont{D.~S.} \bibnamefont{Sanchez}},
  \bibinfo{author}{\bibfnamefont{S.-M.} \bibnamefont{Huang}},
  \bibinfo{author}{\bibfnamefont{I.}~\bibnamefont{Belopolski}},
  \bibinfo{author}{\bibfnamefont{T.-R.} \bibnamefont{Chang}},
  \bibinfo{author}{\bibfnamefont{S.}~\bibnamefont{Zhang}},
  \bibinfo{author}{\bibfnamefont{A.}~\bibnamefont{Bansil}},
  \bibinfo{author}{\bibfnamefont{H.}~\bibnamefont{Lin}}, \bibnamefont{et~al.},
  \bibinfo{journal}{Phys. Rev. Lett.} \textbf{\bibinfo{volume}{119}},
  \bibinfo{pages}{206401} (\bibinfo{year}{2017}).

\bibitem[{\citenamefont{Tang et~al.}(2017)\citenamefont{Tang, Zhou, and
  Zhang}}]{PhysRevLett.119.206402}
\bibinfo{author}{\bibfnamefont{P.}~\bibnamefont{Tang}},
  \bibinfo{author}{\bibfnamefont{Q.}~\bibnamefont{Zhou}}, \bibnamefont{and}
  \bibinfo{author}{\bibfnamefont{S.-C.} \bibnamefont{Zhang}},
  \bibinfo{journal}{Phys. Rev. Lett.} \textbf{\bibinfo{volume}{119}},
  \bibinfo{pages}{206402} (\bibinfo{year}{2017}).

\end{thebibliography}

\end{document}